\documentclass[10pt,reqno,a4paper]{amsart}
\usepackage{amsmath,amssymb,dsfont,graphicx,cite,latexsym,
epsf,cancel,epic,eepic}
\usepackage{amssymb,mathrsfs,txfonts}
\usepackage[colorlinks=false]{hyperref}
\usepackage{mathpazo}
\usepackage{color}
\usepackage{tabularx}
\usepackage{array}
\usepackage{stmaryrd}
\newtheorem{thm}{Theorem}[section]
\newtheorem{prop}[thm]{Proposition}
\newtheorem{cor}[thm]{Corollary}
\newtheorem{lemma}[thm]{Lemma}
\newtheorem{defn}[thm]{Definition}

\theoremstyle{definition}

\newcommand{\R}{\mathbb R}

\newcommand{\Z}{\mathbb Z}
\newcommand{\N}{\mathbb N}

\newcommand{\C}{\mathbb C}

\newcommand{\CP}{\mathbb {CP}}
\newcommand{\x}{\widetilde x}
\newcommand{\y}{\widetilde y}

\newcommand{\T}{\mathrm T}

\newcommand{\Pic}{\mathrm{Pic}}

\newcommand{\X}{{\bf x}}
\newcommand{\XX}{\widetilde{\bf x}}
\renewcommand{\L}{\mathcal L}

\newcommand{\I}{\mathcal I}
\newcommand{\E}{\mathcal E}

\newcommand{\HD}{\mathcal H}
\newcommand{\EE}{\overline{\mathcal E}}
\renewcommand{\H}{\overline H}
\newcommand{\HH}{\widetilde H}
\newcommand{\id}{\mathrm{id}}

\newcommand{\pphi}{\widetilde{\phi}}

\def\ep{\varepsilon}
\def\epsilon{\varepsilon}
\def\beq{\begin{equation}}
\def\eeq{\end{equation}}
\def\bea{\begin{eqnarray}}
\def\eea{\end{eqnarray}}

\title{On the singularity structure of Kahan discretizations of a class of quadratic vector fields}

\author{Ren\'e Zander}

\begin{document}

\maketitle

\begin{center}
{\footnotesize{
Institut f\"ur Mathematik, MA 7-1\\
Technische Universit\"at Berlin, Str. des 17. Juni 136,
10623 Berlin, Germany
}}
\end{center}

\begin{abstract}
\noindent
We discuss the singularity structure of Kahan discretizations of a class of quadratric vector fields and 
provide a classification of the parameter values such that the corresponding Kahan map is integrable, in particular, admits an invariant pencil of elliptic curves.
\end{abstract}

\let\thefootnote\relax\footnote{E-mail: zander@math.tu-berlin.de}

\section{Introduction}

The Kahan discretization scheme was introduced in the unpublished notes \cite{K} as a method applicable to any system of ordinary differential equations in $\R^n$ with a 
quadratic vector field
\begin{equation*}
f(\X)=Q(\X)+B\X+c,\quad \X\in\R^n,
\end{equation*}
where each component of $Q:\R^n \to \R^n$ is a quadratic form, while $B\in\R^{n\times n}$  and
$c \in \R^n$.
Kahan's discretizations reads as
\begin{equation}
\label{KHKdiscretization}
\frac{\XX-\X}{2\epsilon}=Q(\X,\XX)+\frac12B(\X+\XX)+c,
\end{equation}
where 
\begin{equation*}
Q(\X,\XX)=\frac12\left(Q(\X+\XX)-Q(\X)-Q(\XX)\right)
\end{equation*}
is the symmetric bilinear form corresponding to the quadratic form $Q$.
Equation (\ref{KHKdiscretization}) is linear with respect to $\X$ and therefore defines a rational map $\XX=\phi_{\epsilon}(\X)$.
Since equation (\ref{KHKdiscretization}) remains invariant under the interchange
$\X\leftrightarrow\XX$ with the simultaneous sign
inversion $\epsilon\mapsto-\epsilon$, one has the reversibility property 
$\phi_{\epsilon}^{-1}(\X)=\phi_{-\epsilon}(\X)$.
In particular, the map $\phi_{\epsilon}$ is \textit{birational}.

In this paper, we consider the class of two-dimensional quadratic differential equations
\begin{equation}
\begin{pmatrix}
\dot x\\
\dot y
\end{pmatrix}
 =\ell_1^{1-\gamma_1}(x,y)\ell_2^{1-\gamma_2}(x,y)\ell_3^{1-\gamma_3}(x,y)J\nabla H(x,y),\label{nahm_intro}
\end{equation}
where
\begin{equation*}
H(x,y)=\ell_1^{\gamma_1}(x,y)\ell_2^{\gamma_2}(x,y)\ell_3^{\gamma_3}(x,y),
\end{equation*}
and
\begin{equation*}
\ell_i(x,y)=a_ix+b_iy
\end{equation*}
are linear forms, with $a_i,b_i\in\C$, 
$J=\begin{pmatrix}
0&1\\
-1&0
\end{pmatrix}$ 
and $\gamma_1, \gamma_2, \gamma_3\in\R\setminus\{0\}$.

Integrability of the Kahan maps $\phi\colon\C^2\rightarrow\C^2$ has been established for several cases of parameters $(\gamma_1,\gamma_2,\gamma_3)$: 
If $(\gamma_1,\gamma_2,\gamma_3)=(1,1,1)$, then (\ref{nahm_intro}) is a canonical Hamiltonian system on $\R^2$ with homogeneous cubic Hamiltonian. For such systems, a rational integral for the Kahan map $\phi$ has been found in \cite{PPS2,CMOQ13}.
The Kahan maps for the cases $(\gamma_1,\gamma_2,\gamma_3)=(1,1,2)$ and $(\gamma_1,\gamma_2,\gamma_3)=(1,2,3)$ have been treated in 
\cite{PPS2,PZ17,CMMOQ17}.
In all three cases, the level sets of the integral for both the continuous time system and the Kahan discretization have genus $1$.
If $(\gamma_1,\gamma_2,\gamma_3)=(1,1,0)$, then (\ref{nahm_intro}) is a Hamiltonian vector field on $\R^2$ with linear Poisson tensor and homogeneous quadratic Hamiltonian. In this case, a rational integral for the Kahan map $\phi$ has been found in \cite{CMMOQ14}.
The level sets of the integral have genus $0$.

In this paper, we study the singularity structure of the Kahan discretization as a birational quadratic map $\phi\colon\CP^2\rightarrow\CP^2$.
Based on general classification results by Diller and Favre \cite{DF01}, we provide the following classification for the Kahan map $\phi$
of (\ref{nahm_intro}) depending on the values of the parameters $(\gamma_1,\gamma_2,\gamma_3)$:

\begin{thm}
\label{thm:classifiacation_intro}
Let $\phi\colon\CP^2\rightarrow\CP^2$ be the Kahan map of (\ref{nahm_intro}).

The sequence of degrees $d(m)$ of iterates $\phi^m$ grows exponentially, so that the map $\phi$ is non-integrable, except for the following cases:

\begin{enumerate}

\item[(i)] If $(\gamma_1,\gamma_2,\gamma_3)=(1,1,1),(1,1,2),(1,2,3)$, the sequence $d(m)$ of degrees grows quadratically. 
The map $\phi$ admits an invariant pencil of elliptic curves. The degree of a generic curve of the pencil is $3$, $4$, $6$, respectively.

\item[(ii)] If $(\gamma_1,\gamma_2,\gamma_3)=(1,1,0)$ or $(\gamma_1,\gamma_2,\gamma_3)=(\alpha,1,-1)$, $\alpha\in\R\setminus\Z\cup\{0\}$, the sequence of degrees $d(m)$ grows linearly. 
The map $\phi$ admits an invariant pencil of rational curves.

\item[(iii)] If $(\gamma_1,\gamma_2,\gamma_3)=(n,1,-1)$, $n\in\N$, the sequence of degrees $d(m)$ is bounded.

\end{enumerate}

Here, $(\gamma_1,\gamma_2,\gamma_3)$ are fixed up to permutation and multiplication by $\lambda\in\R\setminus\{0\}$.

\end{thm}

Some of the integrable cases are discussed in further detail in sections \ref{sec:Nahm_111}--\ref{sec:Nahm_n1-1}.

\section{Preliminary results}

\subsection{Birational maps of surfaces}

\begin{defn}
Let $\phi$ be a birational map of a smooth projective surface $X$. The dynamical degree of the map $\phi$ is defined as
\begin{equation*}
\lambda_1=\lim\limits_{m\rightarrow\infty}\lVert (\phi^m)^*\rVert^{1/n},
\end{equation*}
where $(\phi^m)^*$ denote the induced pullback maps on the Picard group $\Pic(X)$.
\end{defn}

Diller and Favre provide the following classification for birational maps with $\lambda_1=1$:

\begin{thm}[Diller, Favre \cite{DF01}, Theorem 0.2]\label{thm:DF}
Let $\phi\colon X\rightarrow X$ be a birational map of a smooth projective surface with $\lambda_1=1$.
Up to birational conjugacy, exactly one of the following holds.
\begin{enumerate}

\item[(i)] The sequence $\lVert (\phi^m)^*\rVert$ is bounded, and $\phi^m$ is an automorphism isotopic to the identity for some $m$.

\item[(ii)] The sequence $\lVert (\phi^m)^*\rVert$ grows linearly, and $\phi$ preserves a rational fibration. In this case, $\phi$ cannot 
be conjugated to an automorphism.

\item[(iii)] The sequence $\lVert (\phi^m)^*\rVert$ grows quadratically, and $\phi$ is an automorphism preserving an elliptic fibration.

\end{enumerate}
\end{thm}

One says that $\phi\colon X\rightarrow X$ is \textit{analytically stable} (AS) if $(\phi^*)^m=(\phi^m)^*$ on $\Pic(X)$. This relates the dynamical degree $\lambda_1$ to
the spectral radius of the induced pullback $\phi^*\colon\Pic(X)\rightarrow\Pic(X)$. Equivalently, analytic stability is 
characterized by the condition that there is no curve $V\subset X$ such that $\phi^n(V)\in\I(\phi)$ for some integer $n\geq0$, where $\I(\phi)$ is the indeterminacy set of $\phi$ (see \cite{DF01}, Theorem 1.14).
Therefore, the notion of analytic stability is closely related to \textit{singularity confinement} (see \cite{MWRG19}). 
Indeed, a singularity confinement pattern for a map $\phi\colon X\rightarrow X$ involves a curve $V\subset X$ such that $\phi(V)=P$ is a point (so that $P\in\I(\phi^{-1})$)
and $\phi^{n-1}(P)\in\I(\phi)$, so that $\phi^{n}(P)$ is a curve again for some positive integer $n\in\N$. 
Such a singularity confinement pattern can be resolved by blowing up the orbit of $P$. 
Upon resolving all singularity confinement patterns, one lifts $\phi$ to an AS map $\pphi\colon X'\rightarrow X'$.

Diller and Favre showed that for any birational map $\phi\colon X\rightarrow X$ of a smooth projective surface we can construct
by a finite number of successive blow-ups a surface $X'$ such that 
$\phi$ lifts to an analytically stable birational map $\pphi\colon X'\rightarrow X'$ (see \cite{DF01}, Theorem 0.1).

\subsection{Birational quadratic maps of $\CP^2$}

As shown, e.g., in \cite{CD13}, every quadratic birational map $\phi\colon\CP^2\rightarrow\CP^2$ can be 
represented as $\phi=A_1\circ q_i\circ A_2$, where $A_1, A_2$ are linear projective transformations of $\CP^2$
and $q_i$ is one of the three standard quadratic involutions:
\begin{eqnarray}
&&q_1\colon[x,y,z]\rightarrow [yz,xz,xy],\label{q1}\\
&&q_2\colon[x,y,z]\rightarrow [xz,yz,x^2],\label{q2}\\
&&q_3\colon[x,y,z]\rightarrow [x^2,xy,y^2+xz].\label{q3}
\end{eqnarray}

In these three cases, the indeterminacy set $\I(\phi)$ consists of three, respectively two, one (distinct) singularities. The last two cases correspond to a
coalescence of singularities. Therefore, the first case is the generic one.

In the present work, we only consider the first case: $\phi=A_1\circ q_1\circ A_2$.
In this case, $\I(\phi)=\{B_{+}^{(1)}, B_{+}^{(2)}, B_{+}^{(3)}\}$ consists of three distinct points.
Let $L_{-}^{(i)}$ denote the line through $B_{+}^{(j)}, B_{+}^{(k)}$ (we have $B_{+}^{(i)}=L_{-}^{(j)}\cap L_{-}^{(k)}$).
These lines are exceptional in the sense that they are blown down by $\phi$ to points: $\phi(L_{-}^{(i)})=B_{-}^{(i)}$.
The inverse map is also quadratic with set of indeterminacy points $\I(\phi^{-1})~=~\{B_{-}^{(1)}, B_{-}^{(2)}, B_{-}^{(3)}\}$.

Suppose that the map admits $s$ singularity confinement patterns ($0\leq s\leq3$). That means there are positive integers
$n_1,\dotsc,n_s\in\N$ and $(\sigma_1,\dotsc,\sigma_s)$ such that $\phi^{n_i-1}(B_{-}^{(i)})=B_{+}^{(\sigma_i)}$ for $i=1,\dotsc s$.
We assume that the $n_i$ are taken to be minimal and, for simplicity, we also assume that $\phi^k(B_{-}^{(i)})\neq\phi^l(B_{-}^{(j)})$ for any $k,l\geq0$
and $i\neq j$.
As shown by Bedford and Kim \cite{BK06} one can resolve the singularity confinement patterns by blowing up the finite sequences
$B_{-}^{(i)},\phi(B_{-}^{(i)}),\dotsc,\phi^{n_i-1}(B_{-}^{(i)})$. Those sequences are also called \textit{singular orbits}.
In this paper, we only encounter the situation that the orbits of different $B_{-}^{(i)}$ are disjoint. As shown in \cite{BK06}, one can
adjust the procedure to the more general situation.

On the blow-up surface $X$, the lifted map $\pphi\colon X\rightarrow X$ is AS, and is an automorphism if and only if $s=3$. 
The $s$-tuples $(n_1,\dotsc,n_s)$, $(\sigma_1,\dotsc,\sigma_s)$ are called \textit{orbit data} associated to $\phi$. We say that the map $\phi$ realizes the orbit data $(n_1,\dotsc,n_s)$, $(\sigma_1,\dotsc,\sigma_s)$.

\smallskip

Let $\HD\in\Pic(X)$ be the pullback of the divisor class of a generic line in $\CP^2$. Let $\E_{i,n}\in\Pic(X)$, for $i\leq s$ and $0\leq n\leq n_i-1$, be the divisor class of the exceptional divisor associated to the blow-up of the point $\phi^{n}(B_{-}^{(i)})$.
Then $\HD$ and $\E_{i,n}$ give a basis for $\Pic(X)$, i.e.,
\begin{equation*}
\Pic(X)=\Z\HD\bigoplus\limits_{i=1}^3\bigoplus\limits_{n=0}^{n_i-1}\Z\E_{i,n}
\end{equation*}
that is orthogonal w.r.t. the intersection product, $(\cdot,\cdot)\colon\Pic(X)\times\Pic(X)\rightarrow\Z$, and is normalized by 
$(\HD,\HD)~=~1$ and $(\E_{i,n},\E_{i,n})=-1$. The rank of the Picard group is $\sum n_i+1$.

The induced pullback $\pphi^*\colon\Pic(X)\rightarrow\Pic(X)$ is determined by (see Bedford, Kim, \cite{ BK06} and Diller, \cite{Dil11})
\begin{equation}
\label{pullback}
\begin{array}{llll}
\HD&\mapsto &2\HD-\displaystyle\sum\limits_{j\leq s}\E_{j,n_j-1}, &\vspace{.1truecm}\\
\E_{i,0}&\mapsto&\HD-\displaystyle\sum\limits_{j\leq s\colon\sigma_j\neq i}\E_{j,n_j-1}, & i\leq s,\vspace{.1truecm}\\
\E_{i,n}&\mapsto &\E_{i,n-1}, & i\leq s,\quad 1\leq n\leq n_i-1.
\end{array}
\end{equation}

The induced pushforward $\pphi_*\colon\Pic(X)\rightarrow\Pic(X)$ is determined by
\begin{equation}
\label{pushforward}
\begin{array}{llll}
\HD&\mapsto &2\HD-\displaystyle\sum\limits_{j\leq s}\E_{j,0}, &\vspace{.1truecm}\\
\E_{i,n_i-1}&\mapsto&\HD-\displaystyle\sum\limits_{j\leq s\colon j\neq\sigma_i}\E_{j,0}, & i\leq s,\vspace{.1truecm}\\
\E_{i,n-1}&\mapsto &\E_{i,n}, & i\leq s,\quad 1\leq n\leq n_i-1.
\end{array}
\end{equation}

The maps $\pphi^*, \pphi_*$ are adjoint w.r.t. the intersection product (see \cite{DF01}, Proposition 1.1), i.e., $(\pphi^*A,B)=(A,\pphi_*B)$ for
all $A, B\in\Pic(X)$.

Bedford and Kim have computed the characteristic polynomial $\chi(\lambda)=\det (\pphi^*-\lambda \id)$ explicitly for any given orbit data (see \cite{BK06}, Theorem 3.3).

Let $C(m)=(\pphi^*)^m(\HD)\in\Pic(X)$ be the class of the $m$-th iterate of a generic line.
Set 
\begin{equation}
\label{d(m)}
d(m)=(C(m),\HD),
\end{equation}
so that $d(m)$ is the algebraic degree of the $m$-th iterate of the map $\phi$.
Set 
\begin{equation}
\label{mu(m)}
\mu_i(m+j)=(C(m),\E_{i,j}),\qquad i\leq s,\quad 0\leq j\leq n_i-1.
\end{equation}
The expression on the right-hand side indeed depends on $i$ and $m+j$ only: using that the maps $\pphi^*$, $\pphi_*$ are adjoint w.r.t. the intersection product and the relations (\ref{pushforward}), we find
\begin{equation*}
(C(m),\E_{i,j})=(C(m),\pphi_*\E_{i,j-1})=(\pphi^*C(m),\E_{i,j-1})=(C(m+1),\E_{i,j-1}).
\end{equation*}
In particular, $\mu_i(m)=(C(m),\E_{i,0})$ can be interpreted as the multiplicity of $B_{-}^{(i)}$ 
on the $m$-th iterate of a generic line.

The sequence of degrees $d(m)$ of iterates of the map $\phi$ satisfies a system of linear recurrence relations.

\begin{thm}[Recurrence relations]
\label{thm:recurrence}

Let $\phi$ be a birational map of $\CP^2$ with three distinct indeterminacy points,
and with associated orbit data $(n_1,\dotsc,n_s)$, $(\sigma_1,\dotsc,\sigma_s)$. The degree of iterates $d(m)$ satisfies the system of recurrence relations
\begin{equation}
\label{recurrence}
\left\{ \begin{array}{llll}
d(m+1)&=&2d(m)-\displaystyle\sum\limits_{j\leq s}\mu_j(m),&\vspace{.1truecm}\\
\mu_i(m+n_i)&=&d(m)-\displaystyle\sum\limits_{j\leq s\colon j\neq\sigma_i}\mu_j(m),&  i\leq s,
\end{array} \right.
\end{equation}
with initial conditions $d(0)=1$ and $\mu_i(m)=0$, for $i\leq s$ and $m=0,\dotsc,n_i-1$.

\begin{proof}
With (\ref{d(m)}), (\ref{mu(m)}) we find that
\begin{equation*}
C(m)=d(m)\HD-\sum\limits_{i\leq s}\sum\limits_{j=0}^{n_i-1}\mu_i(m+j)\E_{i,j}.
\end{equation*}
With relations (\ref{pullback}) we compute the pullback
\begin{equation*}
\pphi^*C(m)=d(m)\left(2\HD-\sum\limits_{i\leq s}\E_{i,n_i-1}\right)-\sum\limits_{i\leq s}\left(\sum\limits_{j=1}^{n_i-1}\mu_i(m+j)\E_{i,j-1}+\mu_i(m)\left(\HD-\sum\limits_{j\leq s\colon\sigma_j\neq i}\E_{j,n_j-1}\right)\right).
\end{equation*}
Then we find
\begin{equation*}
\label{rec}
\begin{array}{llll}
(\pphi^*C(m),\HD)&=&2d(m)-\displaystyle\sum\limits_{j\leq s}\mu_j(m),&\vspace{.1truecm}\\
(\pphi^*C(m),\E_{i,n_i-1})&=&d(m)-\displaystyle\sum\limits_{j\leq s\colon j\neq\sigma_i}\mu_j(m),&  i\leq s,\vspace{.1truecm}\\
(\pphi^*C(m),\E_{i,j})&=&\mu_i(m+1+j),& i\leq s,\quad 0\leq j\leq n_i-2.
\end{array}
\end{equation*}
Finally, with $C(m+1)=\pphi^*C(m)$, we obtain the recurrence relations (\ref{recurrence}).
The initial conditions are $d(0)~=~(\HD,\HD)=1$ and $\mu_i(j)=(\HD,\E_{i,j})=0$, for $i\leq s$ and $0\leq j\leq n_i-1$. This proves the claim.
\end{proof}
\end{thm}

\begin{cor}[Generating functions]
\label{cor:generating_functions}
Consider the generating functions $d(z)$, $\mu_i(z)$ for the sequences from theorem \ref{thm:recurrence}.
They are rational functions which can be definded as solutions of the functional equations (\ref{recurrence_gf})
with initial conditions as in theorem \ref{thm:recurrence}.
\begin{equation}
\label{recurrence_gf}
\left\{ \begin{array}{llll}
\dfrac{1}{z}(d(z)-1)&=&2d(z)-\displaystyle\sum\limits_{j\leq s}\mu_j(z),&\vspace{.1truecm}\\
\dfrac{1}{z^{n_i}}\mu_i(z)&=&d(z)-\displaystyle\sum\limits_{j\leq s\colon j\neq\sigma_i}\mu_j(z),&  i\leq s.
\end{array} \right.
\end{equation}
\end{cor}

\section{The $(\gamma_1,\gamma_2,\gamma_3)$-class}

The class of quadratic differential equations we want to consider is a generalization of the two-dimensional reduced Nahm systems introduced in \cite{HMM},
\begin{equation} \label{nahm_HMM}
\left\{ \begin{array}{l}
\dot x = x^2 - y^2, \vspace{.1truecm}\\
\dot y = - 2xy,
\end{array} \right.
\qquad
\left\{ \begin{array}{l}
\dot x = 2 x^2 - 12y^2, \vspace{.1truecm}\\
\dot y = - 6 xy - 4 y^2,
\end{array} \right. \qquad 
\left\{ \begin{array}{l}
\dot x_1 = 2 x^2 - y^2, \vspace{.1truecm}\\
\dot x_2 = - 10 xy + y^2.
\end{array} \right.
\end{equation}
Such systems can be explicitly integrated in terms of elliptic functions and they admit
integrals of motion given respectively by\
\begin{equation*}
H_1(x,y)= \frac{y}{3}(3x^2-y^2),\qquad
H_2(x,y)= y(2x+3y)(x-y)^2,\qquad
H_3(x,y)= \frac{y}{6}(3x-y)^2(4x+y)^3.
\end{equation*}
Note that the curves $\{H_i(x,y)=\lambda\}$ are of genus $1$.
Systems (\ref{nahm_HMM}) have been discussed in \cite{HMM} and
discretized by means of the Kahan method in \cite{PPS2}. Integrability of Kahan discretizations
\begin{equation*}
\left\{ \begin{array}{l}
\x-x = 2\epsilon(\x x-\y y), \vspace{.1truecm}\\
\y-y = -2\epsilon(\x y+x\y),
\end{array} \right.\qquad
\left\{ \begin{array}{l}
\x-x = \epsilon(4\x x-24\y y), \vspace{.1truecm}\\
\y-y = -\epsilon(6\x y+6x\y+8\y y),
\end{array} \right.\qquad
\left\{ \begin{array}{l}
\x-x = \epsilon(4\x x-2\y y), \vspace{.1truecm}\\
\y-y = \epsilon(-10\x y-10x\y+2\y y),
\end{array} \right.
\end{equation*}
was shown in \cite{PPS2}.
They have been studied in the context of minimization of rational elliptic surfaces in \cite{CT13}.
The following generalization of reduced Nahm systems has been introduced in \cite{PZ17,CMMOQ17}:

We use the notation $\X=(x,y)\in\C^2$. Consider the two-dimensional quadratic differential equations
\begin{align}
\label{nahm}
\begin{split}
\dot \X &=\ell_1^{1-\gamma_1}(\X)\ell_2^{1-\gamma_2}(\X)\ell_3^{1-\gamma_3}(\X)J\nabla H(\X),\\
&=\gamma_1\ell_2(\X)\ell_3(\X)J\nabla\ell_1+\gamma_2\ell_1(\X)\ell_3(\X)J\nabla\ell_2+\gamma_3\ell_1(\X)\ell_2(\X)J\nabla\ell_3,
\end{split}
\end{align}
where 
\begin{equation}
\label{nahm_integral}
H(\X)=\ell_1^{\gamma_1}(\X)\ell_2^{\gamma_2}(\X)\ell_3^{\gamma_3}(\X),
\end{equation}
and
\begin{equation*}
\ell_i(x,y)=a_ix+b_iy
\end{equation*}
are linear forms, with $a_i,b_i\in\C$, 
$J=\begin{pmatrix}
0&1\\
-1&0
\end{pmatrix}$ 
and $\gamma_1, \gamma_2, \gamma_3\in\R\setminus\{0\}$. System (\ref{nahm}) has the function (\ref{nahm_integral})
as an integral of motion
and an invariant measure form
\begin{equation}
\label{nahm_measure}
\Omega(\X)=\frac{\mathrm dx\wedge\mathrm dy}{\ell_1(\X)\ell_2(\X)\ell_3(\X)}.
\end{equation}

The Kahan discretization of (\ref{nahm}) reads
\begin{align}
\label{nahm_HK}
\begin{split}
\XX-\X=& \epsilon\gamma_1(\ell_2(\X)\ell_3(\XX)+\ell_2(\XX)\ell_3(\X))J\nabla\ell_1+\epsilon\gamma_2(\ell_1(\X)\ell_3(\XX)+\ell_1(\XX)\ell_3(\X))J\nabla\ell_2\\
&+\epsilon\gamma_3(\ell_1(\X)\ell_2(\XX)+\ell_1(\XX)\ell_2(\X) )J\nabla\ell_3.
\end{split}
\end{align}
It was shown in \cite{PZ17} that the Kahan map admits (\ref{nahm_measure}) as invariant measure form.
Now, multiplying (\ref{nahm_HK}) from the left by the vectors $\nabla\ell_i^{\T}$, $i=1,2,3$, we obtain
\begin{eqnarray}
\ell_1(\XX) -\ell_1(\X)&=&\ep d_{12}\gamma_2(\ell_1(\X)\ell_3(\XX)+\ell_1(\XX)\ell_3(\X))-\ep d_{31}\gamma_3(\ell_1(\X)\ell_2(\XX)+\ell_1(\XX)\ell_2(\X)),
\label{nahm_e1}\\
\ell_2(\XX) -\ell_2(\X)&=&\ep d_{23}\gamma_3(\ell_1(\X)\ell_2(\XX)+\ell_1(\XX)\ell_2(\X))-\ep d_{12}\gamma_1(\ell_2(\X)\ell_3(\XX)+\ell_2(\XX)\ell_3(\X)),
\label{nahm_e2}\\
\ell_3(\XX) -\ell_3(\X)&=&\ep d_{31}\gamma_1(\ell_2(\X)\ell_3(\XX)+\ell_2(\XX)\ell_3(\X))-\ep d_{23}\gamma_2(\ell_1(\X)\ell_3(\XX)+\ell_1(\XX)\ell_3(\X)),
\label{nahm_e3}
\end{eqnarray}
where
\begin{equation*}
d_{ij}=a_ib_j-a_jb_i.
\end{equation*}
From equations (\ref{nahm_e1}--\ref{nahm_e3}) it follows that the Kahan map leaves the lines $\{\ell_i(\X)=0\}$, $i=1,2,3$, invariant.

Explicitly, the Kahan discretization of (\ref{nahm}) as map $\phi_{+}\colon\CP^2\rightarrow\CP^2$ is as follows:
\begin{equation}
\phi_{+}\colon[x,y,z]\longrightarrow[x',y',z']\label{phi+}
\end{equation}
with
\begin{eqnarray}
x'&=&zx+\epsilon A_2(x,y),\label{phi+x}\\
y'&=&zy-\epsilon B_2(x,y),\label{phi+y}\\
z'&=&z^2+z\epsilon C_1(x,y)-2\epsilon^2 C_2(x,y),\label{phi+t}
\end{eqnarray}
with homogeneous polynomials of $\deg\leq2$
\begin{eqnarray*}
A_2(x,y)&=&\sum\limits_{(i,j,k)} \gamma_i\ell_i(x,y)(b_k\ell_j(x,y)+b_j\ell_k(x,y)),\\
B_2(x,y)&=&\sum\limits_{(i,j,k)} \gamma_i\ell_i(x,y)(a_k\ell_j(x,y)+a_j\ell_k(x,y)),\\
C_1(x,y)&=&\sum\limits_{(i,j,k)} \gamma_i(d_{ik}\ell_j(x,y)+d_{ij}\ell_k(x,y)),\\
C_2(x,y)&=&\sum\limits_{(i,j,k)} \gamma_j\gamma_kd_{jk}^2\ell_i^2(x,y),
\end{eqnarray*}
where $\sum\limits_{(i,j,k)}$ denotes the sum over all cyclic permutations of $(i,j,k)$ of $(1,2,3)$.

The inverse $\phi_{-}\colon\CP^2\rightarrow\CP^2$ of the Kahan map (\ref{phi+}) is obtained by replacing $\epsilon$ by $-\epsilon$.

\begin{lemma}
The following identities hold:
\begin{eqnarray}
A_2(-\lambda b_i,\lambda a_i)&=&-b_id_{ij}d_{ki}(\gamma_j+\gamma_k)\lambda^2,\label{A2_ijk}\\ 
B_2(-\lambda b_i,\lambda a_i)&=&-a_id_{ij}d_{ki}(\gamma_j+\gamma_k)\lambda^2,\label{B2_ijk}\\
C_1(-\lambda b_i,\lambda a_i)&=&-d_{ij}d_{ki}(2\gamma_i-\gamma_j-\gamma_k)\lambda,\label{C1_ijk}\\
C_2(-\lambda b_i,\lambda a_i)&=&\gamma_i d_{ij}^2d_{ki}^2(\gamma_j+\gamma_k)\lambda^2,\label{C2_ijk}
\end{eqnarray}
where $(i,j,k)$ is a cyclic permutation of $(1,2,3)$.
\begin{proof}
This is the result of straightforward computations.
\end{proof}
\end{lemma}

In the following, we assume that $d_{12},d_{23},d_{31}\neq0$, i.e., that the lines $\{\ell_i(x,y)=0\}$ are pairwise distinct.
Also, we consider $\C^2$ as affine part of $\CP^2$ consisting of the points $[x,y,z]\in\CP^2$ with $z\neq0$. We identify the point $(x,y)\in\C^2$ 
with the point $[x,y,1]\in\CP^2$.

\begin{prop}
\label{prop:singularities}
The singularities $B_{+}^{(i)}$, $i=1,2,3$, of the Kahan map $\phi_{+}$ and $B_{-}^{(i)}$, $i=1,2,3$, of its inverse $\phi_{-}$ are given by
\begin{equation*}
B^{(i)}_{\pm}=[\pm\frac{b_i}{\epsilon d_{ij}d_{ki}},\mp\frac{a_i}{\epsilon d_{ij}d_{ki}},\gamma_j+\gamma_k],
\end{equation*}
where $(i,j,k)$ is a cyclic permutation of $(1,2,3)$.
Let $\L_{\mp}^{(i)}$ denote the line through the points $B_{\pm}^{(j)}$, $B_{\pm}^{(k)}$. Then we have 
\begin{equation*}
\phi_{\pm}(\L_{\mp}^{(i)})=B_{\mp}^{(i)}.
\end{equation*}
\begin{proof}
Substituting $B_{+}^{(i)}$ into equations (\ref{phi+x})--(\ref{phi+t}) and $B_{-}^{(i)}$ into equations (\ref{phi+x})--(\ref{phi+t}) with $\epsilon$ replaced by $-\epsilon$, and using (\ref{A2_ijk}--\ref{C2_ijk}) the first claim follows immediately.
The second claim is the result of a straightforward (symbolic) computation using Maple.
\end{proof}
\end{prop}

The map $\phi_{+}$ blows down the lines $\L_{-}^{(i)}$ to the points $B_{-}^{(i)}$ and blows up the points $B_{+}^{(i)}$ to the lines $\L_{+}^{(i)}$.

\begin{thm}
\label{thm:orbits} 
\quad
\begin{enumerate}

\item[(i)] Suppose that $n\gamma_i\neq\gamma_j+\gamma_k$, for $0\leq n< N$. Then we have
\begin{equation}
\label{phi^nB-}
\phi_{+}^n(B_{-}^{(i)})=[-\frac{b_i}{\epsilon d_{ij}d_{ki}},\frac{a_i}{\epsilon d_{ij}d_{ki}},-2n\gamma_i+\gamma_j+\gamma_k],\qquad 0\leq n\leq N,
\end{equation}
where $(i,j,k)$ is a cyclic permutation of $(1,2,3)$. In particular, we have
\begin{equation*}
\phi_{+}^{n_i-1}(B_{-}^{(i)})=B_{+}^{(i)}
\end{equation*}
if and only if
\begin{equation}
\label{condition_n_gamma}
(n_i-1)\gamma_i=\gamma_j+\gamma_k,
\end{equation}
for a positive integer $n_i\in\N$. 

\smallskip

\item[(ii)] The only orbit data with exactly three singular orbits that can be realized is $(\sigma_1,\sigma_2,\sigma_3)=(1,2,3)$ and (up to permutation)
\begin{gather*}
(n_1,n_2,n_3)=(3,3,3)\quad\text{if and only if}\quad (\gamma_1,\gamma_2,\gamma_3)=\lambda(1,1,1),\\
(n_1,n_2,n_3)=(4,4,2)\quad\text{if and only if}\quad (\gamma_1,\gamma_2,\gamma_3)=\lambda(1,1,2),\\
(n_1,n_2,n_3)=(6,3,2)\quad\text{if and only if}\quad (\gamma_1,\gamma_2,\gamma_3)=\lambda(1,2,3),
\end{gather*}
for $\lambda\in\R\setminus\{0\}$.

\smallskip

\item[(iii)] The only orbit data with exactly two singular orbits that can be realized is $(\sigma_1,\sigma_2)=(1,2)$ and 
\begin{equation*}
(n_1,n_2)\in N_2=\N^2\setminus\{(3,3),(2,4),(4,2),(4,4),(2,3),(3,2),(2,6),(6,2),(3,6),(6,3)\}
\end{equation*}
if and only if
\begin{equation*}
(\gamma_1,\gamma_2,\gamma_3)=\lambda(n_2,n_1,n_1n_2-n_1-n_2),
\end{equation*}
for $\lambda\in\R\setminus\{0\}$.

\smallskip

\item[(iv)] The only orbit data with exactly one singular orbit that can be realized is $\sigma_1=1$ and $n_1\in\N$ arbitrary.

\end{enumerate}

\begin{proof}
\quad

\noindent (i) We show (\ref{phi^nB-}) by induction on $n$. For $n=0$ the claim is true by Proposition \ref{prop:singularities}.
In the induction step (from $n<N$ to $n+1$) with (\ref{phi+x}--\ref{phi+t}) and (\ref{A2_ijk}--\ref{C2_ijk}) we find that
\begin{eqnarray*}
x'&=&-\frac{2(-n\gamma_i+\gamma_j+\gamma_k)b_i}{\epsilon d_{ij}d_{ki}},\\
y'&=&\frac{2(-n\gamma_i+\gamma_j+\gamma_k)a_i}{\epsilon d_{ij}d_{ki}},\\
z'&=&2(-n\gamma_i+\gamma_j+\gamma_k)(-2(n+1)\gamma_i+\gamma_j+\gamma_k).
\end{eqnarray*}
Since $n\gamma_i\neq\gamma_j+\gamma_k$, we find that
\begin{equation*}
\phi_{+}(\phi_{+}^n(B_{-}^{(i)}))=[-\frac{b_i}{\epsilon d_{ij}d_{ki}},\frac{a_i}{\epsilon d_{ij}d_{ki}},-2(n+1)\gamma_i+\gamma_j+\gamma_k].
\end{equation*}
This proves the claim.

\smallskip

\noindent (ii) From conditions (\ref{condition_n_gamma}), for $i=1,2,3$, we obtain the linear system
\begin{equation}
\label{eqs_s=3}
\begin{pmatrix}
n_1-1&-1&-1\\
-1&n_2-1&-1\\
-1&-1&n_3-1
\end{pmatrix}
\begin{pmatrix}
\gamma_1\\
\gamma_2\\
\gamma_3
\end{pmatrix}
=
\begin{pmatrix}
0\\
0\\
0
\end{pmatrix}.
\end{equation}
This system has nontrivial solutions if and only if $n_1n_2n_3-n_1n_2-n_1n_3-n_2n_3=0$. This yields the proof.

\smallskip

\noindent (iii) 
From conditions (\ref{condition_n_gamma}), for $i=1,2$, we obtain the linear system
\begin{equation}
\label{eqs_s=2}
\begin{pmatrix}
n_1-1&-1&-1\\
-1&n_2-1&-1
\end{pmatrix}
\begin{pmatrix}
\gamma_1\\
\gamma_2\\
\gamma_3
\end{pmatrix}
=
\begin{pmatrix}
0\\
0
\end{pmatrix}.
\end{equation}
Note that we have to exclude those values $(n_1,n_2)\in\N^2$ for which the solutions $(\gamma_1,\gamma_2,\gamma_3)$ correspond to orbit data with three singular orbits.
This yields the proof.

\smallskip

\noindent (iv)
From conditions (\ref{condition_n_gamma}), for $i=1$, we obtain the linear equation
\begin{equation}
\label{eqs_s=1}
\begin{pmatrix}
n_1-1&-1&-1
\end{pmatrix}
\begin{pmatrix}
\gamma_1\\
\gamma_2\\
\gamma_3
\end{pmatrix}
=
0.
\end{equation}
This yields the proof.

\end{proof}
\end{thm}

We arrive at the following classification result (compare Theorem \ref{thm:classifiacation_intro}):

\begin{thm}
\label{thm:classification}

The sequence of degrees $d(m)$ of iterates $\phi_{+}^m$ grows exponentially, so that the map $\phi_{+}$ is non-integrable, except for the following cases:

\begin{enumerate}

\item[(i)] If $(\gamma_1,\gamma_2,\gamma_3)=(1,1,1),(1,1,2),(1,2,3)$, the sequence $d(m)$ of degrees grows quadratically. 
The map $\phi_{+}$ admits an invariant pencil of elliptic curves. The degree of a generic curve of the pencil is $3$, $4$, $6$, respectively.

\item[(ii)] If $(\gamma_1,\gamma_2,\gamma_3)=(1,1,0)$ or $(\gamma_1,\gamma_2,\gamma_3)=(\alpha,1,-1)$, $\alpha\in\R\setminus\Z\cup\{0\}$, the sequence of degrees $d(m)$ grows linearly. 
The map $\phi_{+}$ admits an invariant pencil of rational curves.

\item[(iii)] If $(\gamma_1,\gamma_2,\gamma_3)=(n,1,-1)$, $n\in\N$, the sequence of degrees $d(m)$ is bounded.

\end{enumerate}

Here, $(\gamma_1,\gamma_2,\gamma_3)$ are fixed up to permutation and multiplication by $\lambda\in\R\setminus\{0\}$.

\begin{proof}

We distinguish the number of singular orbits $s=0,1,2,3$ of the map $\phi_{+}$.

\smallskip

$s=3$.
If $(\gamma_1,\gamma_2,\gamma_3)=(1,1,1),(1,1,2),(1,2,3)$, the generating functions of the sequences of degrees are given by (\ref{gf_111}), (\ref{gf_112}) and (\ref{gf_123}), respectively. The sequences $d(m)$ grow quadratically. The invariant pencils of elliptic curves are given by (\ref{111_curves}), (\ref{112_curves}) and (\ref{123_curves}), 
respectively.
By Theorem \ref{thm:orbits} these are the only cases with three singular orbits.

\smallskip

$s=2$. 
If $(\gamma_1,\gamma_2,\gamma_3)=(1,1,0)$, the sequence of degrees is given by (\ref{gf_110}). The sequence $d(m)$ grows linearly.
The invariant pencil of rational curves is given by (\ref{110_curves}).
If $(\gamma_1,\gamma_2,\gamma_3)=(n,1,-1)$, $n\in\N$, the generating function of the sequence of degrees is given by (\ref{gf_n1-1}). The sequence $d(m)$ is bounded.
By Theorem \ref{thm:orbits} all other cases with two singular orbits have orbit data $(\sigma_1,\sigma_2)=(1,2)$, $(n_1,n_2)=(2+i,2+j)$ with $i+j>2$. With Theorem 3.3 in \cite{BK06} and Theorem 5.1 in \cite{BK04} it follows that in those cases $\lambda_1>1$, i.e., the sequence $d(m)$ grows exponentially.

\smallskip

$s=1$.
If $(\gamma_1,\gamma_2,\gamma_3)=(\alpha,1,-1)$, $\alpha\in\R\setminus\Z\cup\{0\}$, by Theorem \ref{thm:orbits} and (\ref{eqs_s=1}) we have the orbit data $\sigma_1=1$, $n_1=1$. With Theorem \ref{thm:recurrence} we find that the sequence $d(m)$ grows linearly.
The claim about the existence of an invariant pencil of rational curves follows from Theorem \ref{thm:DF}.
With (\ref{eqs_s=1}) we find that all other cases with one singular orbit have orbit data $\sigma_1=1$, $n_1>1$. With Theorem 3.3 in \cite{BK06} and Theorem 5.1 in \cite{BK04} it follows that in those cases $\lambda_1>1$, i.e., the sequence $d(m)$ grows exponentially.

\smallskip

$s=0$. We have $\lambda_1=2$. The sequence $d(m)$ grows exponentially.

\end{proof}

\end{thm}

\section{The case $(\gamma_1,\gamma_2,\gamma_3)=(1,1,1)$}\label{sec:Nahm_111}

By Theorem \ref{thm:orbits} this case corresponds to the orbit data $(n_1,n_2,n_3)=(3,3,3)$, $(\sigma_1,\sigma_2,\sigma_3)=(1,2,3)$.

In this case, we consider the Kahan map $\phi_+\colon\C^2\rightarrow\C^2$ corresponding to a quadratic vector field of the form
\begin{equation*}
\dot{\X}=J\nabla H(\X),\quad H(\X)=\ell_1(\X)\ell_2(\X)\ell_3(\X).
\end{equation*}

The Kahan map $\phi_+\colon\C^2\rightarrow\C^2$ admits an integral of motion (see \cite{CMOQ13,PSS19}): 
\begin{equation}
\label{111_integral}
\HH(\X)=\dfrac{H(\X)}{P(\X)},
\end{equation}
where
\begin{eqnarray*}
P(\X)&=&1+4\epsilon^2((d_1d_3-d_2^2)x^2+(d_1d_4-d_2d_3)xy+(d_2d_4-d_3^2)y^2),
\end{eqnarray*}
with $d_1=3a_1a_2a_3$, $d_2=a_1a_2b_3+a_1a_3b_2+a_2a_3b_1$, $d_3=a_3b_1b_2+a_2b_1b_3+a_1b_2b_3$, $d_4=3b_1b_2b_3$.

The geometry of the Kahan discretization has been studied in \cite{PSS19}.
The phase space of $\phi_{+}\colon\C^2\rightarrow\C^2$ is foliated by the one-parameter family (pencil) of invariant curves
\begin{equation*}
\E_{\lambda}=\left\{(x,y)\in\C^2\colon H(x,y)-\lambda P(x,y)=0\right\}.
\end{equation*}
We consider $\C^2$ as an affine part of $\CP^2$ consisting of the points $[x,y,z]\in\CP^2$ with $z\neq0$.
We define the projective curves $\EE_{\lambda}$ as projective completion on $\E_{\lambda}$:
\begin{equation}
\label{111_curves}
\EE_{\lambda}=\left\{[x,y,z]\in\CP^2\colon H(x,y)-\lambda z\overline{P}(x,y,z)=0\right\},
\end{equation}
where we set
\begin{equation*}
\overline{P}(x,y,z)=z^2P_2(x/z,y/z).
\end{equation*}
(We have $\H(x,y,z)=z^3H(x/z,y/z)=H(x,y)$ since $H(x,y)$ is homogeneous of degree three.)
The pencil contains two reducible curves
\begin{equation*}
\EE_0=\{[x,y,z]\in\CP^2\colon H(x,y)=0\}
\end{equation*}
consisting of the lines $\{\ell_i(x,y)=0\}$, $i=1,2,3$, and 
\begin{equation*}
\EE_{\infty}=\{[x,y,z]\in\CP^2\colon z\overline{P}(x,y,z)=0\}
\end{equation*}
consisting of the conic $\{\overline{P}(x,y,z)=0\}$ and the line at infinity $\{z=0\}$.
All curves $\EE_{\lambda}$ pass through the set of base points which is defined as $\EE_0\cap\EE_{\infty}$. 
According to the B\'ezout theorem, there are $9$ base points, counted with multiplicities.

\begin{prop}
The $9$ base points are given by:

two finite base points of multiplicity $1$ on each of the lines $\ell_i=0$, $i=1,2,3$:
\begin{equation}
B^{(i)}_{\pm}=(\pm\frac{b_i}{2\epsilon d_{ij}d_{ki}},\mp\frac{a_i}{2\epsilon d_{ij}d_{ki}}),\label{111_B123}
\end{equation}

one base point of multiplicity $1$ at infinity on each of the lines $\ell_i=0$, $i=1,2,3$:
\begin{equation}
F^{(i)}=[b_i,-a_i,0].\label{111_F123}
\end{equation}

The singular orbits of the map are as follows:
\begin{eqnarray}
\begin{aligned}\label{111_orbits}
&\L_{-}^{(1)}\longrightarrow B_{-}^{(1)}\longrightarrow F^{(1)}\longrightarrow B_{+}^{(1)}\longrightarrow \L_{+}^{(1)},\\
&\L_{-}^{(2)}\longrightarrow B_{-}^{(2)}\longrightarrow F^{(2)}\longrightarrow B_{+}^{(2)}\longrightarrow \L_{+}^{(2)},\\
&\L_{-}^{(3)}\longrightarrow B_{-}^{(3)}\longrightarrow F^{(3)}\longrightarrow B_{+}^{(3)}\longrightarrow \L_{+}^{(3)},
\end{aligned}
\end{eqnarray}
where $\L_{\mp}^{(i)}$ denotes the line through the points $B_{\pm}^{(j)}$, $B_{\pm}^{(k)}$.

\begin{proof}
The singular orbits (\ref{111_orbits}) are a consequence of Proposition \ref{prop:singularities} and Theorem \ref{thm:orbits}. It can be verified by straightforward computations that the points (\ref{111_B123})--(\ref{111_F123}) are base points of the pencil of invariant curves $\EE_{\lambda}$.
\end{proof}
\end{prop}

\begin{figure}[htbp]
\includegraphics[scale=0.7]{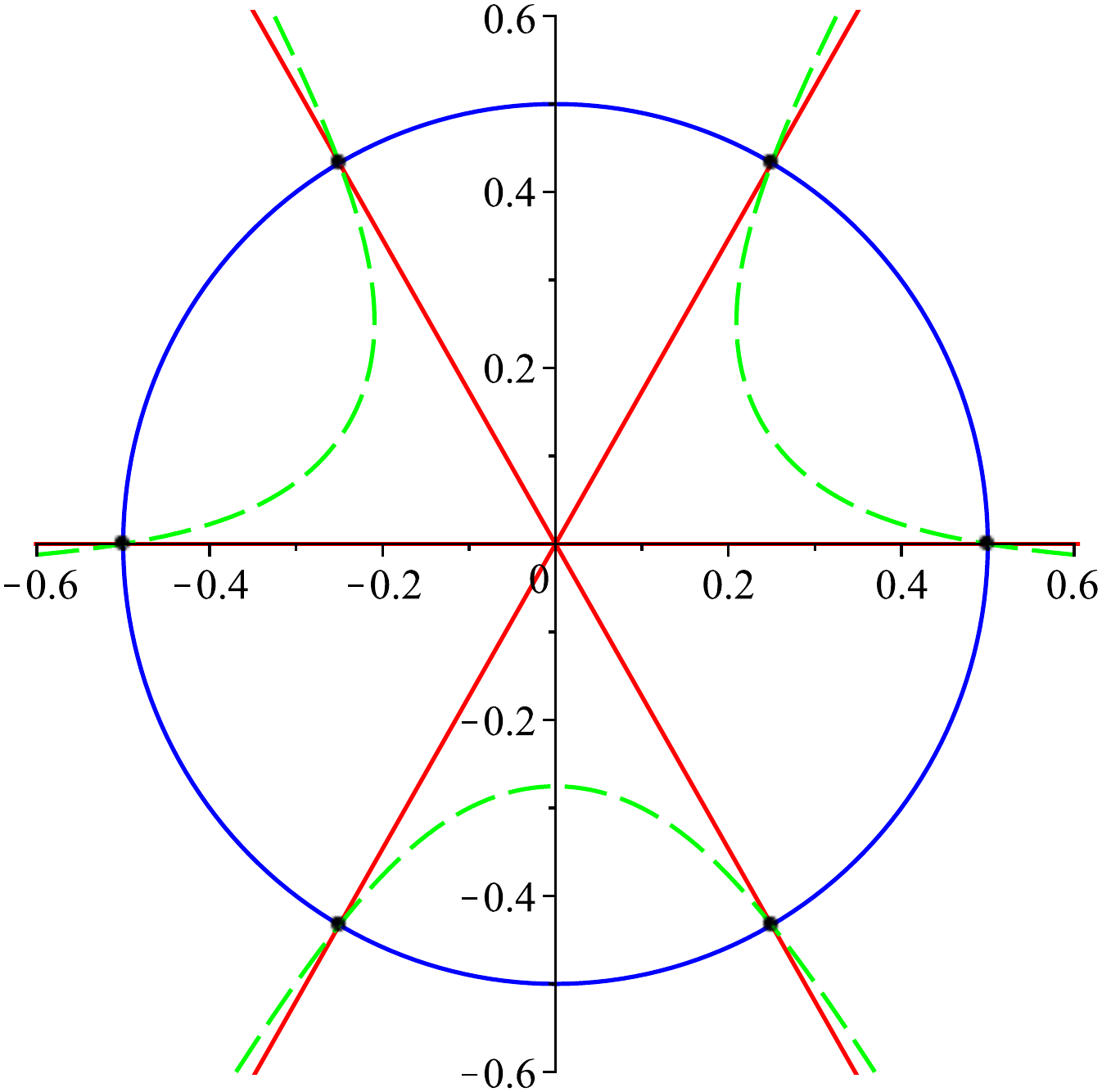}
\put(-25,120){$x$}
\put(-125,255){$y$}
\put(-25,145){$B_{-}^{(1)}$}
\put(-260,145){$B_{+}^{(1)}$}
\put(-215,230){$B_{-}^{(2)}$}
\put(-70,35){$B_{+}^{(2)}$}
\put(-215,35){$B_{-}^{(3)}$}
\put(-70,230){$B_{+}^{(3)}$}
\caption{The curves $\EE_0$, $\EE_{\infty}$, $\EE_{0.01}$ in resp. red, blue and green for $H(x,y)=H_1(x,y)$, $\epsilon=1$.}
\end{figure}

\subsection{Lifting the map to a surface automorphism}
We blow up the plane $\CP^2$ at the nine base points $B_{-}^{(i)},F^{(i)},B_{+}^{(i)}$ ($i=1,2,3$) and 
denote the corresponding exceptional divisors by $E_{i,0},E_{i,1},E_{i,2}$ ($i=1,2,3$). The resulting blow-up surface is
denoted by $X$. On this surface $\phi_{+}$ is lifted to an automorphism $\pphi_{+}$ acting on the exceptional divisors according to the scheme (compare with (\ref{111_orbits}))
\begin{eqnarray*}
\begin{aligned}
&\widetilde{\L}_{-}^{(1)}\longrightarrow E_{1,0}\longrightarrow E_{1,1}\longrightarrow E_{1,2}\longrightarrow \widetilde{\L}_{+}^{(1)},\\
&\widetilde{\L}_{-}^{(2)}\longrightarrow E_{2,0}\longrightarrow E_{2,1}\longrightarrow E_{2,2}\longrightarrow \widetilde{\L}_{+}^{(2)},\\
&\widetilde{\L}_{-}^{(3)}\longrightarrow E_{3,0}\longrightarrow E_{3,1}\longrightarrow E_{3,2}\longrightarrow \widetilde{\L}_{+}^{(3)},
\end{aligned}
\end{eqnarray*}
where $\widetilde{\L}_{\pm}^{(i)}$ denotes the proper transform of the line $\L_{\pm}^{(i)}$.

We compute the induced pullback map on the Picard group $\pphi_{+}^*\colon\Pic(X)\rightarrow\Pic(X)$.
Let $\HD\in\Pic(X)$ be the pullback of the class of a generic line in $\CP^2$.
Let $\E_{i,n}\in\Pic(X)$, for $i\leq 3$ and $0\leq n\leq 2$, be the class of $E_{i,n}$.
Then the Picard group is
\begin{equation*}
\Pic(X)=\Z\HD\bigoplus\limits_{i=1}^3\bigoplus\limits_{n=0}^{2}\Z\E_{i,n},
\end{equation*}
The rank of the Picard group is $10$.
The induced pullback $\pphi_{+}^*\colon\Pic(X)\rightarrow\Pic(X)$ is determined by (\ref{pullback}).

With Theorem \ref{thm:recurrence} we arrive at the system of recurrence relations for the degree $d(m)$:
\begin{equation*}
\left\{ \begin{array}{l}
d(m+1)=2d(m)-\mu_1(m)-\mu_2(m)-\mu_3(m), \vspace{.1truecm}\\
\mu_1(m+3)=d(m)-\mu_2(m)-\mu_3(m), \vspace{.1truecm}\\
\mu_2(m+3)=d(m)-\mu_1(m)-\mu_3(m), \vspace{.1truecm}\\
\mu_3(m+3)=d(m)-\mu_1(m)-\mu_2(m),
\end{array} \right.
\end{equation*}
with initial conditions $d(0)=1$, $\mu_i(m)=0$, for $m=0,\dotsc,2$, $i=1,2,3$.
The generating functions of the solution to this system of recurrence relations are given by:
\begin{eqnarray}
d(z)&=&-\dfrac{2z^3+1}{(z+1)(z-1)^3},\label{gf_111}\\
\mu_i(z)&=&-\dfrac{z^3}{(z+1)(z-1)^3},\qquad i=1,2,3.\notag
\end{eqnarray}

The sequence $d(m)$ grows quadratically.

\section{The case $(\gamma_1,\gamma_2,\gamma_3)=(1,1,2)$}\label{sec:Nahm_112}

By Theorem \ref{thm:orbits} this case corresponds to the orbit data $(n_1,n_2,n_3)=(4,4,2)$, $(\sigma_1,\sigma_2,\sigma_3)=(1,2,3)$.

In this case, we consider the Kahan map $\phi_+\colon\C^2\rightarrow\C^2$ corresponding to a quadratic vector field of the form
\begin{equation*}
\dot{\X}=\frac{1}{\ell_3(\X)}J\nabla H(\X),\quad H(\X)=\ell_1(\X)\ell_2(\X)\ell_3^2(\X).
\end{equation*}

The Kahan map $\phi_+\colon\C^2\rightarrow\C^2$ admits an integral of motion (see \cite{CMMOQ17,PZ17}): 
\begin{equation}
\label{112_integral}
\HH(\X)=\dfrac{H(\X)}{P_1^{(1)}(\X)P_1^{(2)}(\X)P_2(\X)},
\end{equation}
where
\begin{eqnarray*}
P_1^{(1)}(\X)&=&1+\epsilon\left(d_{23}\ell_1(\X)-d_{31}\ell_2(\X)\right),\\
P_1^{(2)}(\X)&=&1-\epsilon\left(d_{23}\ell_1(\X)-d_{31}\ell_2(\X)\right),\\
P_2(\X)&=&1-\epsilon^2\left(9d_{12}^2\ell_3^2(\X)-4d_{23}d_{31}\ell_1(\X)\ell_2(\X)\right).
\end{eqnarray*}

The phase space of $\phi_{+}\colon\C^2\rightarrow\C^2$ is foliated by the one-parameter family (pencil) of invariant curves
\begin{equation*}
\E_{\lambda}=\left\{(x,y)\in\C^2\colon H(x,y)-\lambda P_1^{(1)}(x,y)P_1^{(2)}(x,y)P_2(x,y)=0\right\}.
\end{equation*}
We define the projective curves $\EE_{\lambda}$ as projective completion on $\E_{\lambda}$:
\begin{equation}
\label{112_curves}
\EE_{\lambda}=\left\{[x,y,z]\in\CP^2\colon H(x,y)-\lambda \overline{P}_1^{(1)}(x,y,z)\overline{P}_1^{(2)}(x,y,z)\overline{P}_2(x,y,z)=0\right\},
\end{equation}
where we set
\begin{equation*}
\overline{P}_1^{(i)}(x,y,z)=zP_1^{(i)}(x/z,y/z),\quad\,i=1,2,\qquad \overline{P}_2(x,y,z)=z^2P_2(x/z,y/z).
\end{equation*}
(We have $\H(x,y,z)=z^4H(x/z,y/z)=H(x,y)$ since $H(x,y)$ is homogeneous of degree four.)
The pencil contains two reducible curves
\begin{equation*}
\EE_0=\{[x,y,z]\in\CP^2\colon H(x,y)=0\}
\end{equation*}
consisting of the lines $\{\ell_i(x,y)=0\}$, $i=1,2,3$, with multiplicities $1,1,2$, and 
\begin{equation*}
\EE_{\infty}=\{[x,y,z]\in\CP^2\colon \overline{P}_1^{(1)}(x,y,z)\overline{P}_1^{(2)}(x,y,z)\overline{P}_2(x,y,z)=0\}
\end{equation*}
consisting of the two lines $\{\overline{P}_1^{(i)}(x,y,z)=0\}$, $i=1,2$, and the conic $\{\overline{P}_2(x,y,z)=0\}$.
All curves $\EE_{\lambda}$ pass through the set of base points which is defined as $\EE_0\cap\EE_{\infty}$.

\begin{prop}
The $10$ (distinct) base points are given by:

four base points of multiplicity $1$ on each of the lines $\ell_i=0$, $i=1,2$:
\begin{align}
&B^{(i)}_{\pm}=(\pm\frac{b_i}{3\epsilon d_{ij}d_{ki}},\mp\frac{a_i}{3\epsilon d_{ij}d_{ki}}),\label{112_B12}\\
&C^{(i)}_{\pm}=(\pm\frac{b_i}{\epsilon d_{ij}d_{ki}},\mp\frac{a_i}{\epsilon d_{ij}d_{ki}}),\label{112_C12}
\end{align}

two base points of multiplicity $2$ on the line $\ell_3=0$:
\begin{equation}
B^{(3)}_{\pm}=(\pm\frac{b_3}{2\epsilon d_{23}d_{31}},\mp\frac{a_3}{2\epsilon d_{23}d_{31}}).\label{112_B3}
\end{equation}

The singular orbits of the map are as follows:
\begin{eqnarray}
\begin{aligned}\label{112_orbits}
&\L_{-}^{(1)}\longrightarrow B_{-}^{(1)}\longrightarrow C_{-}^{(1)}\longrightarrow C_{+}^{(1)}\longrightarrow B_{+}^{(1)}\longrightarrow \L_{+}^{(1)},\\
&\L_{-}^{(2)}\longrightarrow B_{-}^{(2)}\longrightarrow C_{-}^{(2)}\longrightarrow C_{+}^{(2)}\longrightarrow B_{+}^{(2)}\longrightarrow \L_{+}^{(2)},\\
&\L_{-}^{(3)}\longrightarrow B_{-}^{(3)}\longrightarrow B_{+}^{(3)}\longrightarrow \L_{+}^{(3)},
\end{aligned}
\end{eqnarray}
where $\L_{\mp}^{(i)}$ denotes the line through the points $B_{\pm}^{(j)}$, $B_{\pm}^{(k)}$.

\begin{proof}
The singular orbits (\ref{112_orbits}) are a consequence of Proposition \ref{prop:singularities} and Theorem \ref{thm:orbits}. It can be verified by straightforward computations that the points (\ref{112_B12})--(\ref{112_B3}) are base points of the pencil of invariant curves $\EE_{\lambda}$.
\end{proof}
\end{prop}

According to the B\'ezout theorem, there are $16$ base points, counted with multiplicities. This number
is obtained by 
\begin{equation*}
\sum\limits_{P\in\EE_0\cap\EE_{\infty}}(\mathrm{mult}(P))^2=8\cdot1+2\cdot4,
\end{equation*}
where $\mathrm{mult}(P)$ denotes the multiplicity of the base point $P$.

\begin{figure}[htbp]
\includegraphics[scale=0.7]{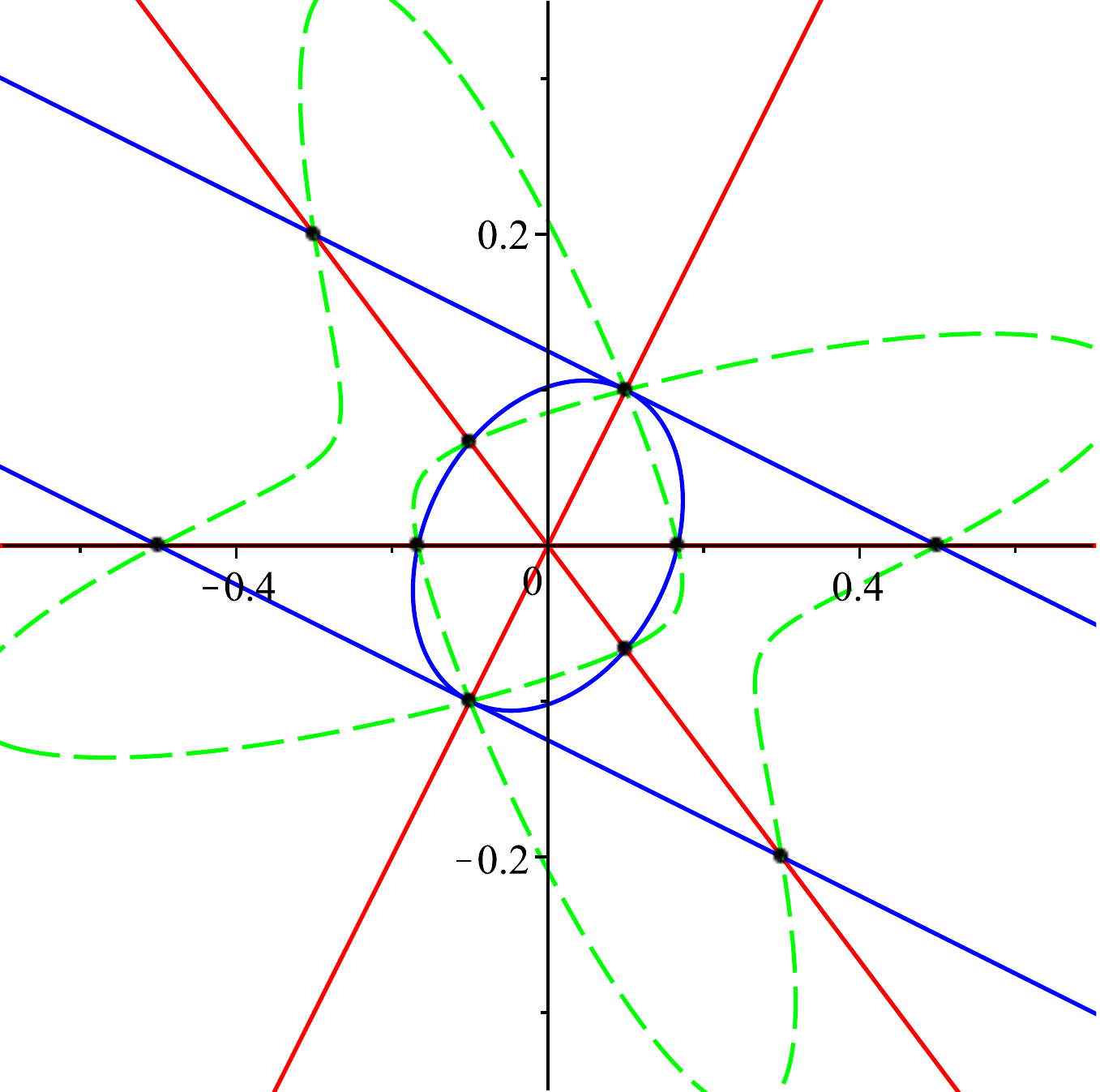}
\put(-5,120){$x$}
\put(-125,255){$y$}
\put(-100,145){$B_{-}^{(1)}$}
\put(-45,145){$C_{-}^{(1)}$}
\put(-240,145){$C_{+}^{(1)}$}
\put(-190,145){$B_{+}^{(1)}$}
\put(-160,175){$B_{-}^{(2)}$}
\put(-190,215){$C_{-}^{(2)}$}
\put(-75,60){$C_{+}^{(2)}$}
\put(-110,105){$B_{+}^{(2)}$}
\put(-180,80){$B_{-}^{(3)}$}
\put(-105,185){$B_{+}^{(3)}$}
\caption{The curves $\EE_0$, $\EE_{\infty}$, $\EE_{0.001}$ in resp. red, blue and green for $H(x,y)=H_2(x,y)$, $\epsilon=1$.}
\end{figure}

\subsection{Lifting the map to a surface automorphism}
We blow up the plane $\CP^2$ at the ten base points $B_{-}^{(i)}, B_{+}^{(i)}$ ($i=1,2,3$) and $C_{-}^{(i)}, C_{+}^{(i)}$ ($i=1,2$) and 
denote the corresponding exceptional divisors by $E_{i,0}\dotsc E_{i,n_i-1}$ ($i=1,2,3$). The resulting blow-up surface is
denoted by $X$. On this surface $\phi_{+}$ is lifted to an automorphism $\pphi_{+}$ acting on the exceptional divisors according to the scheme (compare with (\ref{112_orbits}))
\begin{eqnarray*}
\begin{aligned}
&\widetilde{\L}_{-}^{(1)}\longrightarrow E_{1,0}\longrightarrow E_{1,1}\longrightarrow E_{1,2}\longrightarrow E_{1,3}\longrightarrow \widetilde{\L}_{+}^{(1)},\\
&\widetilde{\L}_{-}^{(2)}\longrightarrow E_{2,0}\longrightarrow E_{2,1}\longrightarrow E_{2,2}\longrightarrow E_{2,3}\longrightarrow \widetilde{\L}_{+}^{(2)},\\
&\widetilde{\L}_{-}^{(3)}\longrightarrow E_{3,0}\longrightarrow E_{3,1}\longrightarrow \widetilde{\L}_{+}^{(3)},
\end{aligned}
\end{eqnarray*}
where $\widetilde{\L}_{\pm}^{(i)}$ denotes the proper transform of the line $\L_{\pm}^{(i)}$.

We compute the induced pullback map on the Picard group $\pphi_{+}^*\colon\Pic(X)\rightarrow\Pic(X)$. 
Let $\HD\in\Pic(X)$ be the pullback of the class of a generic line in $\CP^2$.
Let $\E_{i,n}\in\Pic(X)$, for $i\leq 3$ and $0\leq n\leq n_i-1$, be the class of $E_{i,n}$.
Then the Picard group is
\begin{equation*}
\Pic(X)=\Z\HD\bigoplus\limits_{i=1}^3\bigoplus\limits_{n=0}^{n_i-1}\Z\E_{i,n}.
\end{equation*}
The rank of the Picard group is $11$. 
The induced pullback $\pphi_{+}^*\colon\Pic(X)\rightarrow\Pic(X)$ is determined by (\ref{pullback}).

With Theorem \ref{thm:recurrence} we arrive at the system of recurrence relations for the degree $d(m)$:
\begin{equation*}
\left\{ \begin{array}{l}
d(m+1)=2d(m)-\mu_1(m)-\mu_2(m)-\mu_3(m), \vspace{.1truecm}\\
\mu_1(m+4)=d(m)-\mu_2(m)-\mu_3(m), \vspace{.1truecm}\\
\mu_2(m+4)=d(m)-\mu_1(m)-\mu_3(m), \vspace{.1truecm}\\
\mu_3(m+2)=d(m)-\mu_1(m)-\mu_2(m),
\end{array} \right.
\end{equation*}
with initial conditions $d(0)=1$, $\mu_i(m)=0$, for $n=0,\dotsc,3$, $i=1,2$, and $\mu_3(m)=0$, for $m=0,1$.
The generating functions of the solution to this system of recurrence relations are given by:
\begin{eqnarray}
d(z)&=&-\dfrac{2z^4+z^2+1}{(z^2+z+1)(z-1)^3},\label{gf_112}\\
\mu_i(z)&=&-\dfrac{z^4}{(z^2+z+1)(z-1)^3},\qquad i=1,2,\notag\\
\mu_3(z)&=&-\dfrac{z^2(z^2+1)}{(z^2+z+1)(z-1)^3}.\notag
\end{eqnarray}

The sequence $d(m)$ grows quadratically.

\section{The case $(\gamma_1,\gamma_2,\gamma_3)=(1,2,3)$}\label{sec:Nahm_123}

By Theorem \ref{thm:orbits} this case corresponds to the orbit data $(n_1,n_2,n_3)=(6,3,2)$, $(\sigma_1,\sigma_2,\sigma_3)=(1,2,3)$.

In this case, we consider the Kahan map $\phi_+\colon\C^2\rightarrow\C^2$ corresponding to a quadratic vector field of the form
\begin{equation*}
\dot{\X}=\frac{1}{\ell_2(\X)\ell_3^2(\X)}J\nabla H(\X),\quad H(\X)=\ell_1(\X)\ell_2^2(\X)\ell_3^3(\X).
\end{equation*}

The Kahan map $\phi_+\colon\C^2\rightarrow\C^2$ admits an integral of motion (see \cite{CMMOQ17,PZ17}):
\begin{equation}
\label{123_integral}
\HH(\X)=\dfrac{H(\X)}{P_1^{(1)}(\X)P_1^{(2)}(\X)P_1^{(3)}(\X)P_1^{(4)}(\X)P_2(\X)},
\end{equation}
where
\begin{eqnarray*}
P_1^{(1)}(\X)&=&1+3\epsilon d_{31}\ell_2(\X),\\
P_1^{(2)}(\X)&=&1-3\epsilon d_{31}\ell_2(\X),\\
P_1^{(3)}(\X)&=&1+\epsilon \left(3d_{23}\ell_1(\X)-d_{12}\ell_3(\X)\right),\\
P_1^{(4)}(\X)&=&1-\epsilon \left(3d_{23}\ell_1(\X)-d_{12}\ell_3(\X)\right),\\
P_2(\X)&=&1-\epsilon^2\left(9d_{31}^2\ell_2^2(\X)+16d_{12}^2\ell_3^2(\X)\right).
\end{eqnarray*}

The phase space of $\phi_{+}\colon\C^2\rightarrow\C^2$ is foliated by the one-parameter family (pencil) of invariant curves
\begin{equation*}
\E_{\lambda}=\left\{(x,y)\in\C^2\colon H(x,y)-\lambda P_2(x,y)\prod_{i=1}^4P_1^{(i)}(x,y)=0\right\}.
\end{equation*}
We define the projective curves $\EE_{\lambda}$ as projective completion on $\E_{\lambda}$:
\begin{equation}
\label{123_curves}
\EE_{\lambda}=\left\{[x,y,z]\in\CP^2\colon H(x,y)-\lambda \overline{P}_2(x,y,z)\prod_{i=1}^4\overline{P}_1^{(i)}(x,y,z)=0\right\},
\end{equation}
where we set
\begin{equation*}
\overline{P}_1^{(i)}(x,y,z)=zP_1^{(i)}(x/z,y/z),\quad\,i=1,\dotsc,4,\qquad \overline{P}_2(x,y,z)=z^2P_2(x/z,y/z).
\end{equation*}
The pencil contains two reducible curves
\begin{equation*}
\EE_0=\{[x,y,z]\in\CP^2\colon H(x,y)=0\}
\end{equation*}
consisting of the lines $\{\ell_i(x,y)=0\}$, $i=1,2,3$, with multiplicities $1,2,3$, and  
\begin{equation*}
\EE_{\infty}=\{[x,y,z]\in\CP^2\colon \overline{P}_2(x,y,z)\prod_{i=1}^4\overline{P}_1^{(i)}(x,y,z)=0\}
\end{equation*}
consisting of the four lines $\{\overline{P}_1^{(i)}(x,y,z)=0\}$, $i=1,\dotsc,4$, and the conic $\{\overline{P}_2(x,y,z)=0\}$.
All curves $\EE_{\lambda}$ pass through the set of base points which is defined as $\EE_0\cap\EE_{\infty}$.

\begin{prop}
The 11 (distinct) base points are given by:

six finite base points of multiplicity $1$ on the line $\ell_1=0$:
\begin{gather}
B_{\pm}^{(1)}=(\pm\frac{b_1}{5\epsilon d_{12}d_{31}},\mp\frac{a_1}{5\epsilon d_{12}d_{31}}),\label{123_B1}\\
C_{\pm}^{(1)}=(\pm\frac{b_1}{3\epsilon d_{12}d_{31}},\mp\frac{a_1}{3\epsilon d_{12}d_{31}}),\quad
D_{\pm}^{(1)}=(\pm\frac{b_1}{\epsilon d_{12}d_{31}},\mp\frac{a_1}{\epsilon d_{12}d_{31}}),\label{123_C1}
\end{gather}

two finite base points of multiplicity $2$ on the line $\ell_2=0$:
\begin{equation}
B_{\pm}^{(2)}=(\pm\frac{b_2}{4\epsilon d_{12}d_{23}},\mp\frac{a_2}{4\epsilon d_{12}d_{23}}),\label{123_B2}
\end{equation}

one base point of multiplicity $2$ at infinity one the line $\ell_2=0$:
\begin{equation}
F^{(2)}=[b_2,-a_2,0],\label{123_F}
\end{equation}

two finite base points of multiplicity $3$ on the line $\ell_3=0$:
\begin{equation}
B_{\pm}^{(3)}=(\pm\frac{b_3}{3\epsilon d_{23}d_{31}},\mp\frac{a_3}{3\epsilon d_{23}d_{31}}).\label{123_B3}
\end{equation}

The singular orbits of the map are as follows:
\begin{eqnarray}
\begin{aligned}\label{123_orbits}
&\L_{-}^{(1)}\longrightarrow B_{-}^{(1)}\longrightarrow C_{-}^{(1)}\longrightarrow D_{-}^{(1)}\longrightarrow D_{+}^{(1)}\longrightarrow C_{+}^{(1)}\longrightarrow B_{+}^{(1)}\longrightarrow \L_{+}^{(1)},\\
&\L_{-}^{(2)}\longrightarrow B_{-}^{(2)}\longrightarrow F^{(2)}\longrightarrow B_{+}^{(2)}\longrightarrow \L_{+}^{(2)},\\
&\L_{-}^{(3)}\longrightarrow B_{-}^{(3)}\longrightarrow B_{+}^{(3)}\longrightarrow \L_{+}^{(3)},
\end{aligned}
\end{eqnarray}
where $\L_{\mp}^{(i)}$ denotes the line through the points $B_{\pm}^{(j)}$, $B_{\pm}^{(k)}$.

\begin{proof}
The singular orbits (\ref{123_orbits}) are a consequence of Proposition \ref{prop:singularities} and Theorem \ref{thm:orbits}. It can be verified by straightforward computations that the points (\ref{123_B1})--(\ref{123_B3}) are base points of the pencil of invariant curves $\EE_{\lambda}$.
\end{proof}
\end{prop}

According to the B\'ezout theorem, there are $36$ base points, counted with multiplicities.
This number is obtained by 
\begin{equation*}
\sum\limits_{P\in\EE_0\cap\EE_{\infty}}(\mathrm{mult}(P))^2=6\cdot1+3\cdot4+2\cdot9.
\end{equation*}

\begin{figure}[htbp]
\includegraphics[scale=0.7]{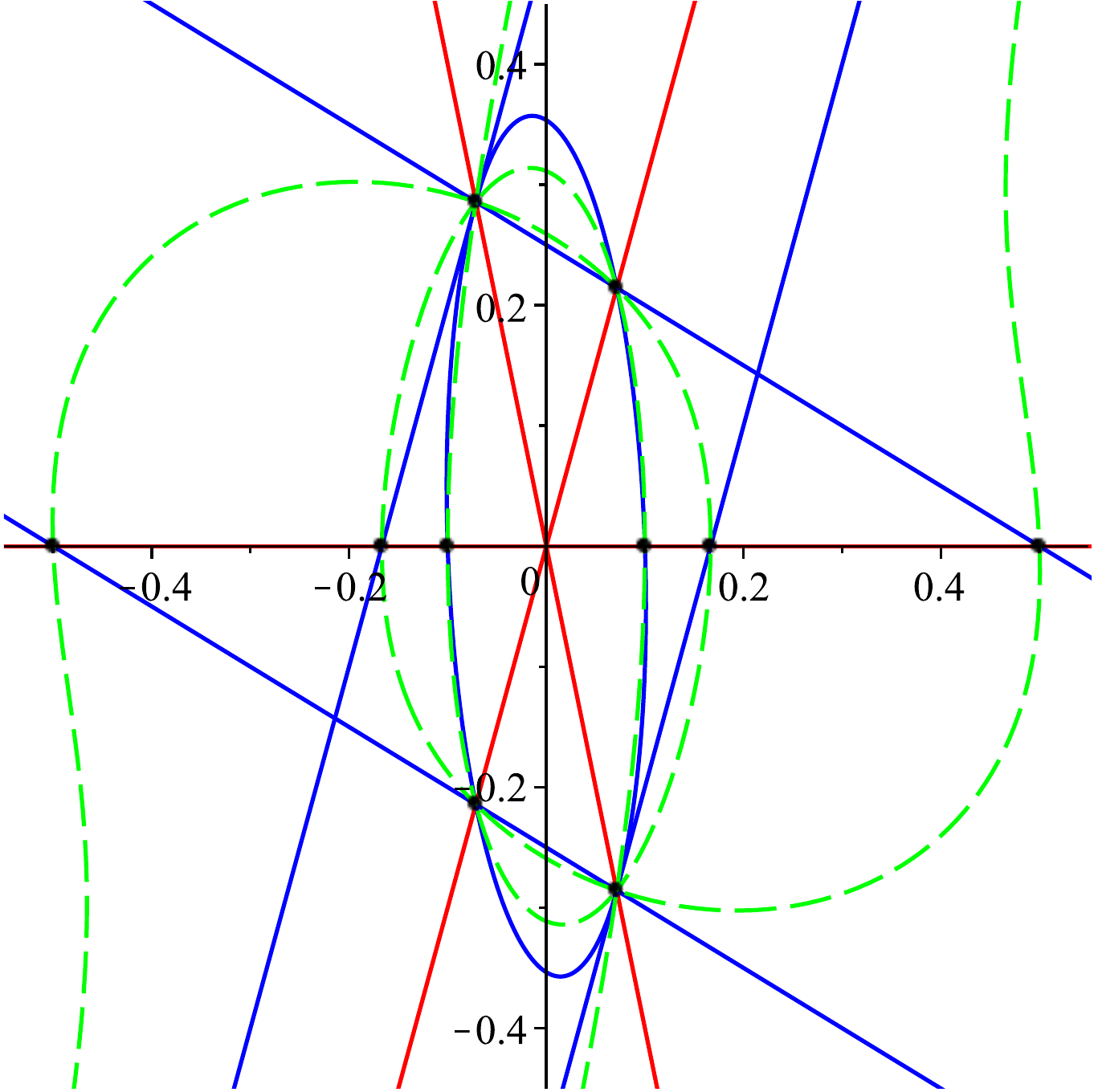}
\put(-5,120){$x$}
\put(-125,255){$y$}
\put(-130,145){$B_{-}^{(1)}$}
\put(-90,145){$C_{-}^{(1)}$}
\put(-10,145){$D_{-}^{(1)}$}
\put(-160,145){$B_{+}^{(1)}$}
\put(-195,145){$C_{+}^{(1)}$}
\put(-260,145){$D_{+}^{(1)}$}
\put(-175,60){$B_{-}^{(2)}$}
\put(-115,205){$B_{+}^{(2)}$}
\put(-180,215){$B_{-}^{(3)}$}
\put(-115,55){$B_{+}^{(3)}$}
\caption{The curves $\EE_0$, $\EE_{\infty}$, $\EE_{-0.002}$ in resp. red, blue and green for $H(x,y)=H_3(x,y)$, $\epsilon=1$.}
\end{figure}

\subsection{Lifting the map to a surface automorphism}
We blow up the plane $\CP^2$ at the eleven base points $B_{-}^{(i)}, B_{+}^{(i)}$ ($i=1,2,3$) and $C_{-}^{(1)}, C_{+}^{(1)}$, $D_{-}^{(1)}, D_{+}^{(1)}$ and $F^{(2)}$ and denote the corresponding exceptional divisors by $E_{i,0}\dotsc E_{i,n_i-1}$ ($i=1,2,3$). The resulting blow-up surface is
denoted by $X$. On this surface $\phi_{+}$ is lifted to an automorphism $\pphi_{+}$ acting on the exceptional divisors according to the scheme (compare with (\ref{123_orbits}))
\begin{eqnarray*}
\begin{aligned}
&\widetilde{\L}_{-}^{(1)}\longrightarrow E_{1,0}\longrightarrow E_{1,1}\longrightarrow E_{1,2}\longrightarrow E_{1,3}\longrightarrow E_{1,4}\longrightarrow E_{1,5}\longrightarrow \widetilde{\L}_{+}^{(1)},\\
&\widetilde{\L}_{-}^{(2)}\longrightarrow E_{2,0}\longrightarrow E_{2,1}\longrightarrow E_{2,2}\longrightarrow \widetilde{\L}_{+}^{(2)},\\
&\widetilde{\L}_{-}^{(3)}\longrightarrow E_{3,0}\longrightarrow E_{3,1}\longrightarrow \widetilde{\L}_{+}^{(3)},
\end{aligned}
\end{eqnarray*}
where $\widetilde{\L}_{\pm}^{(i)}$ denotes the proper transform of the line $\L_{\pm}^{(i)}$.

We compute the induced pullback map on the Picard group $\pphi_{+}^*\colon\Pic(X)\rightarrow\Pic(X)$.
Let $\HD\in\Pic(X)$ be the pullback of the class of a generic line in $\CP^2$.
Let $\E_{i,n}\in\Pic(X)$, for $i\leq 3$ and $0\leq n\leq n_i-1$, be the class of $E_{i,n}$.
Then the Picard group is
\begin{equation*}
\Pic(X)=\Z\HD\bigoplus\limits_{i=1}^3\bigoplus\limits_{n=0}^{n_i-1}\Z\E_{i,n}.
\end{equation*}
The rank of the Picard group is $12$.
The induced pullback $\pphi_{+}^*\colon\Pic(X)\rightarrow\Pic(X)$ is determined by (\ref{pullback}).

With Theorem \ref{thm:recurrence} we arrive at the system of recurrence relations for the degree $d(m)$:
\begin{equation*}
\left\{ \begin{array}{l}
d(m+1)=2d(m)-\mu_1(m)-\mu_2(m)-\mu_3(m), \vspace{.1truecm}\\
\mu_1(m+6)=d(m)-\mu_2(m)-\mu_3(m), \vspace{.1truecm}\\
\mu_2(m+3)=d(m)-\mu_1(m)-\mu_3(m), \vspace{.1truecm}\\
\mu_3(m+2)=d(m)-\mu_1(m)-\mu_2(m),
\end{array} \right.
\end{equation*}
with initial conditions $d(0)=1$, $\mu_1(m)=0$, for $m=0,\dotsc,5$, $\mu_2(m)=0$, for $m=0,1,2$, and $\mu_3(m)=0$, for $m=0,1$.
The generating functions of the solution to this system of recurrence relations are given by:
\begin{eqnarray}
d(z)&=&-\dfrac{2z^6+z^4+z^3+z^2+1}{(z^4+z^3+z^2+z+1)(z-1)^3},\label{gf_123}\\
\mu_1(z)&=&-\dfrac{z^6}{(z^4+z^3+z^2+z+1)(z-1)^3},\notag\\
\mu_2(z)&=&-\dfrac{z^3(z+1)(z^2-z+1)}{(z^4+z^3+z^2+z+1)(z-1)^3},\notag\\
\mu_3(z)&=&-\dfrac{z^2(z^2+z+1)(z^2-z+1)}{(z^4+z^3+z^2+z+1)(z-1)^3}.\notag
\end{eqnarray}

The sequence $d(m)$ grows quadratically.

\section{The case $(\gamma_1,\gamma_2,\gamma_3)=(1,1,0)$}\label{sec:Nahm_110}

By Theorem \ref{thm:orbits} this case corresponds to the orbit data $(n_1,n_2)=(2,2)$, $(\sigma_1,\sigma_2)=(1,2)$.

In this case, we consider the Kahan map $\phi_+\colon\C^2\rightarrow\C^2$ corresponding to a quadratic vector field of the form
\begin{equation*}
\dot{\X}=\ell_3(\X)J\nabla H(\X),\quad H(\X)=\ell_1(\X)\ell_2(\X).
\end{equation*}

For $\ell_1(\X)=x+y$, $\ell_2(\X)=x-y$, $\ell_3(\X)=x$ the vector field reads
\begin{equation*}
\left\{ \begin{array}{l}
\dot x = -2xy, \vspace{.1truecm}\\
\dot y = -2x^2,
\end{array} \right.
\end{equation*}
and the Kahan discretization (\ref{nahm_HK}) reads
\begin{equation*}
\left\{ \begin{array}{l}
\x-x = -2\epsilon(\x y+x\y), \vspace{.1truecm}\\
\y-y = -4\epsilon\x x.
\end{array} \right.
\end{equation*}

The Kahan map $\phi_{+}\colon\C^2\rightarrow\C^2$ admits an integral of motion (see \cite{CMMOQ14,KCMMOQ19}):
\begin{equation}
\HH(\X)=\dfrac{\ell_1(\X)\ell_2(\X)}{P_1^{(1)}(\X)P_1^{(2)}(\X)},
\end{equation}
where
\begin{eqnarray*}
P_1^{(1)}(\X)&=&1+\epsilon d_{12}\ell_3(\X),\\
P_1^{(2)}(\X)&=&1-\epsilon d_{12}\ell_3(\X).
\end{eqnarray*}

The phase space of $\phi_{+}\colon\C^2\rightarrow\C^2$ is foliated by the one-parameter family (pencil) of invariant curves
\begin{equation*}
\E_{\lambda}=\left\{(x,y)\in\C^2\colon H(x,y)-\lambda P_1^{(1)}(x,y)P_1^{(2)}(x,y)=0\right\}.
\end{equation*}
We define the projective curves $\EE_{\lambda}$ as projective completion on $\E_{\lambda}$:
\begin{equation}
\label{110_curves}
\EE_{\lambda}=\left\{[x,y,z]\in\CP^2\colon H(x,y)-\lambda \overline{P}_1^{(1)}(x,y,z)\overline{P}_1^{(2)}(x,y,z)=0\right\},
\end{equation}
where we set
\begin{equation*}
\overline{P}_1^{(i)}(x,y,z)=zP_1^{(i)}(x/z,y/z),\,\text{for}\,i=1,2.
\end{equation*}
(We have $\H(x,y,z)=z^2H(x/z,y/z)=H(x,y)$ since $H(x,y)$ is homogeneous of degree two.)
The pencil contains two reducible curves
\begin{equation*}
\EE_0=\{[x,y,z]\in\CP^2\colon H(x,y)=0\}
\end{equation*}
consisting of the lines $\{\ell_i(x,y)=0\}$, $i=1,2$, and
\begin{equation*}
\EE_{\infty}=\{[x,y,z]\in\CP^2\colon \overline{P}_1^{(1)}(x,y,z)\overline{P}_1^{(2)}(x,y,z)=0\}
\end{equation*}
consisting of the two lines $\{\overline{P}_1^{(i)}(x,y,z)=0\}$, $i=1,2$.
All curves $\EE_{\lambda}$ pass through the set of base points which is defined as $\EE_0\cap\EE_{\infty}$. 
According to the B\'ezout theorem, there are four base points, counted with multiplicities.

\begin{prop}
The four finite base points are given by (see \cite{PS18}):

two base points of multiplicity $1$ on each of the lines $\ell_i=0$, $i=1,2$:
\begin{equation}
B_{\pm}^{(i)}=(\pm\frac{b_i}{\epsilon d_{ij}d_{ki}},\mp\frac{a_i}{\epsilon d_{ij}d_{ki}}).\label{110_B12}
\end{equation}

The singular orbits of the map are as follows:
\begin{eqnarray}
\label{110_orbits}
\begin{aligned}
&\L_{-}^{(1)}\longrightarrow B_{-}^{(1)}\longrightarrow B_{+}^{(1)}\longrightarrow \L_{+}^{(1)},\\
&\L_{-}^{(2)}\longrightarrow B_{-}^{(2)}\longrightarrow B_{+}^{(2)}\longrightarrow \L_{+}^{(2)},
\end{aligned}
\end{eqnarray}
where $\L_{\mp}^{(i)}$ denotes the line through the points $B_{\pm}^{(j)}$, $B_{\pm}^{(k)}$.

\begin{proof}
The singular orbits are a consequence of Proposition \ref{prop:singularities} and Theorem \ref{thm:orbits}. It can be verified by straightforward computations that the points (\ref{110_B12}) are base points of the pencil of invariant curves $\EE_{\lambda}$.
\end{proof}
\end{prop}

With (\ref{phi^nB-}) we see that the point $B_{-}^{(3)}$ is a fixed point of $\phi_{+}$ while $B_{+}^{(3)}$ is a fixed point of $\phi_{-}$. Therefore,
they participate in patterns
\begin{eqnarray*}
&\L_{-}^{(3)}\longrightarrow B_{-}^{(3)}\circlearrowleft,\\
&\circlearrowright B_{+}^{(3)}\longrightarrow \L_{+}^{(3)},
\end{eqnarray*}
which do not qualify as singularity confinement patterns \cite{MWRG19,TEGORS03} and need not be blown up.

\begin{figure}[htbp]
\includegraphics[scale=0.7]{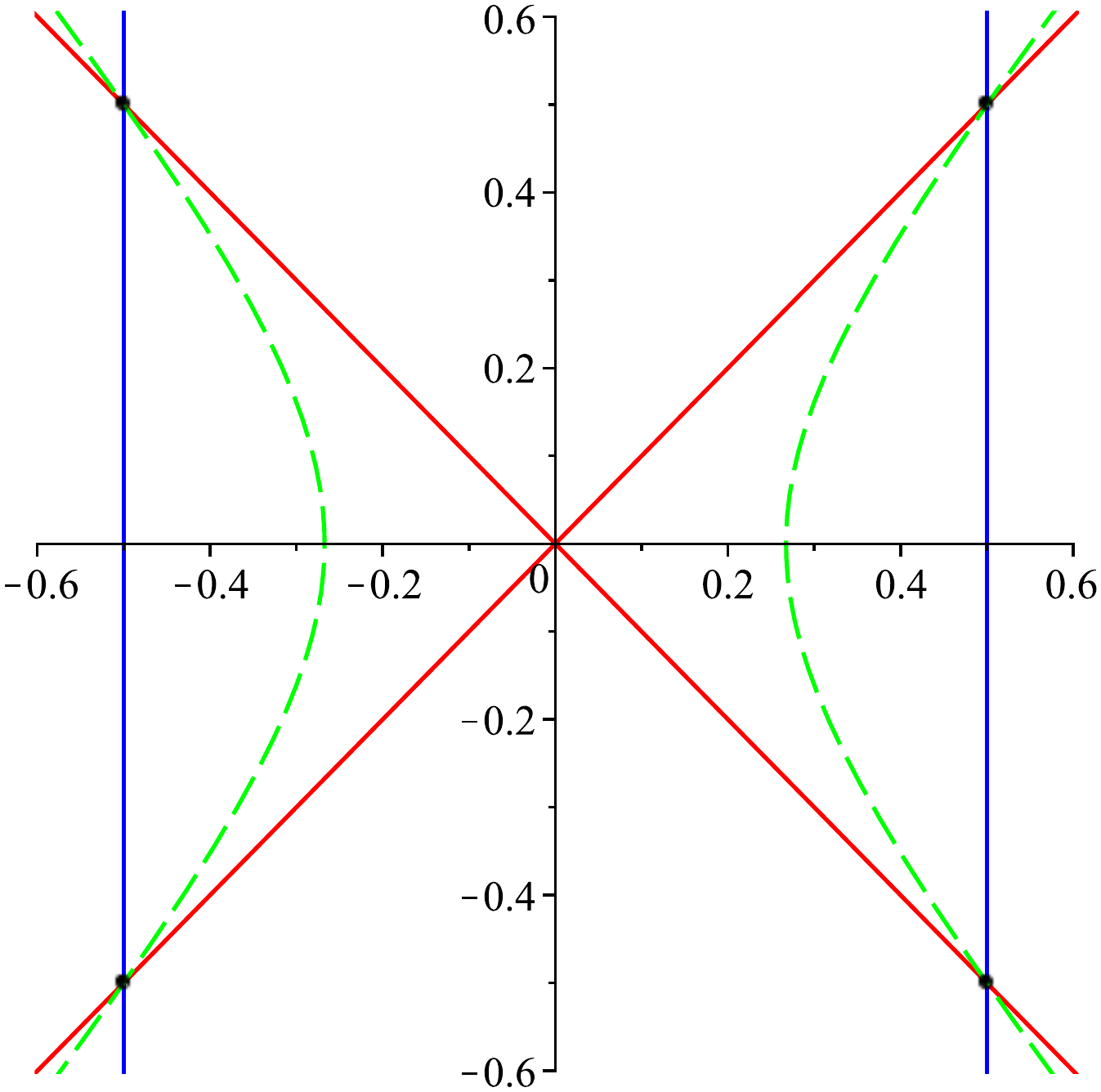}
\put(-25,120){$x$}
\put(-125,255){$y$}
\put(-50,15){$B_{-}^{(1)}$}
\put(-235,245){$B_{+}^{(1)}$}
\put(-235,15){$B_{-}^{(2)}$}
\put(-50,245){$B_{+}^{(2)}$}
\caption{The curves $\EE_0$, $\EE_{\infty}$, $\EE_{0.1}$ in resp. red, blue and green for $\ell_1(x,y)=x+y$, $\ell_2(x,y)=x-y$, $\ell_3(x,y)=x$ and $\epsilon=1$.}
\end{figure}

\subsection{Lifting the map to an analytically stable map}
We blow up the plane $\CP^2$ at the four base points $B_{-}^{(i)}, B_{+}^{(i)}$ ($i=1,2$) and denote the corresponding exceptional divisors by $E_{i,0},E_{i,1}$ ($i=1,2$). The resulting blow-up surface is
denoted by $X$. On this surface $\phi_{+}$ is lifted to an analytically stable map $\pphi_{+}$ acting on the exceptional divisors according to the scheme (compare with (\ref{110_orbits}))
\begin{eqnarray*}
\begin{aligned}
&\widetilde{\L}_{-}^{(1)}\longrightarrow E_{1,0}\longrightarrow E_{1,1}\longrightarrow \widetilde{\L}_{+}^{(1)},\\
&\widetilde{\L}_{-}^{(2)}\longrightarrow E_{2,0}\longrightarrow E_{2,1}\longrightarrow \widetilde{\L}_{+}^{(2)},
\end{aligned}
\end{eqnarray*}
where $\widetilde{\L}_{\pm}^{(i)}$ denotes the proper transform of the line $\L_{\pm}^{(i)}$.

We compute the induced pullback map on the Picard group $\pphi_{+}^*\colon\Pic(X)\rightarrow\Pic(X)$. 
Let $\HD\in\Pic(X)$ be the pullback of the class of a generic line in $\CP^2$.
Let $\E_{i,n}\in\Pic(X)$, for $i=1,2$ and $n=0,1$, be the class of $E_{i,n}$.
Then the Picard group is
\begin{equation*}
\Pic(X)=\Z\HD\bigoplus\limits_{i=1}^2\bigoplus\limits_{n=0}^{1}\Z\E_{i,n}.
\end{equation*}
The rank of the Picard group is $5$.
The induced pullback $\pphi_{+}^*\colon\Pic(X)\rightarrow\Pic(X)$ is determined by (\ref{pullback}).

With Theorem \ref{thm:recurrence} we arrive at the system of recurrence relations for the degree $d(m)$:
\begin{equation*}
\left\{ \begin{array}{l}
d(m+1)=2d(m)-\mu_1(m)-\mu_2(m), \vspace{.1truecm}\\
\mu_1(m+2)=d(m)-\mu_2(m), \vspace{.1truecm}\\
\mu_2(m+2)=d(m)-\mu_1(m),
\end{array} \right.
\end{equation*}
with initial conditions $d(0)=1$, $\mu_1(m)=0$, for $m=0,1$, and $\mu_2(m)=0$, for $m=0,1$.
The solution to this system of recurrence relations is given by:
\begin{eqnarray}
d(m)&=&2m,\label{gf_110}\\
\mu_i(m)&=&m-1,\qquad i=1,2.\notag
\end{eqnarray}

The sequence $d(m)$ grows linearly.

\section{The case $(\gamma_1,\gamma_2,\gamma_3)=(n,1,-1)$}\label{sec:Nahm_n1-1}

By Theorem \ref{thm:orbits} this case corresponds to the orbit data $(n_1,n_2)=(1,n)$, $(\sigma_1,\sigma_2)=(1,2)$.

In this case, we consider the Kahan map $\phi_+\colon\C^2\rightarrow\C^2$ corresponding to a quadratic vector field of the form
\begin{equation*}
\dot{\X}=\frac{\ell_3^2(\X)}{\ell_1^{n-1}(\X)}J\nabla H(\X),\quad H(\X)=\frac{\ell_1^n(\X)\ell_2(\X)}{\ell_3(\X)}.
\end{equation*}
The case $n=1$ has been studied in \cite{PZ17}.

For $\ell_1(\X)=x$, $\ell_2(\X)=x+y$, $\ell_3(\X)=x-y$ the vector field reads
\begin{equation*}
\left\{ \begin{array}{l}
\dot x = 2x^2, \vspace{.1truecm}\\
\dot y = -nx^2+ny^2+2xy,
\end{array} \right.
\end{equation*}
and the Kahan discretization (\ref{nahm_HK}) reads
\begin{equation*}
\left\{ \begin{array}{l}
\x-x = 4\epsilon\x x, \vspace{.1truecm}\\
\y-y = 2\epsilon(-n\x x+n\y y+\x y+x\y).
\end{array} \right.
\end{equation*}

\begin{prop}
The Kahan map $\phi_{+}\colon\C^2\rightarrow\C^2$ admits an integral of motion
\begin{equation}
\HH(\X)=\dfrac{H(\X)}{P(\X)},
\end{equation}
where
\begin{equation}
P(\X)=\prod\limits_{k\in I}(\epsilon d_{23}k\ell_1(\X)+1)(\epsilon d_{23} k\ell_1(\X)-1),
\end{equation}
for $I=\{1,3,5,\dotsc,n-1\}$ if $n$ is even and $I=\{2,4,6,\dotsc,n-1\}$ if $n$ is odd.
\begin{proof}
Note that the following identity holds:
\begin{equation}
\label{e1}
-d_{12}\ell_3(\X)-d_{31}\ell_2(\X)=d_{23}\ell_1(\X).
\end{equation}
Then, using (\ref{e1}), from equation (\ref{nahm_e1}) it follows that
\begin{equation}
\label{e2}
\ell_1(\XX)=\dfrac{\ell_1(\X)}{2\epsilon d_{23}\ell_1(\X)+1}.
\end{equation}
Moreover, multiplying (\ref{nahm_e2}) by $\ell_3(\X)$ and (\ref{nahm_e3}) by $\ell_2(\X)$ and then subtracting the second equation from the first 
equation and again applying (\ref{e1}), we arrive at
\begin{equation}
\label{e3}
\dfrac{\ell_2(\XX)}{\ell_3(\XX)}=-\dfrac{\ell_2(\X)(\epsilon d_{23}(n+1)\ell_1(\X)+1)}{\ell_3(\X)(\epsilon d_{23}(n-1)\ell_1(\X)-1)}.
\end{equation}
On the other hand, from (\ref{e2}) it follows that
\begin{equation}
\label{e4}
\epsilon d_{23}k\ell_1(\XX)\pm1=\dfrac{\epsilon d_{23}(k\pm2)\ell_1(\X)\pm1}{2\epsilon d_{23}\ell_1(\X)+1},
\end{equation}
and therefore, with $h_{\pm}^k(\X)=(\epsilon d_{23}k\ell_1(\X)\pm1)$, we find 
\begin{equation*}
\dfrac{P(\XX)}{P(\X)}=\dfrac{h_-^{-1}(\X)h_-^1(\X)\dotsb h_-^{n-3}(\X)\cdot h_+^3(\X)h_+^5(\X)\dotsb h_+^{n+1}(\X)}
{(h_+^2(\X))^n\cdot h_-^1(\X)h_-^3(\X)\dotsb h_-^{n-1}(\X)\cdot h_+^1(\X)h_+^3(\X)\dotsb h_+^{n-1}(\X)}
=-\dfrac{h_+^{n+1}(\X)}{(h_+^2(\X))^nh_-^{n-1}(\X)},
\end{equation*}
if $n$ is even, and 
\begin{equation*}
\dfrac{P(\XX)}{P(\X)}=\dfrac{h_-^0(\X)h_-^2(\X)\dotsb h_-^{n-3}(\X)\cdot h_+^4(\X)h_+^6(\X)\dotsb h_+^{n+1}(\X)}
{(h_+^2(\X))^{n-1}\cdot h_-^2(\X)h_-^4(\X)\dotsb h_-^{n-1}(\X)\cdot h_+^2(\X)h_+^4(\X)\dotsb h_+^{n-1}(\X)}
=-\dfrac{h_+^{n+1}(\X)}{(h_+^2(\X))^nh_-^{n-1}(\X)},
\end{equation*}
if $n$ is odd.
This proves the claim.
\end{proof}
\end{prop}

With Theorem \ref{thm:recurrence} we arrive at the system of recurrence relations for the degree $d(m)$:
\begin{equation*}
\left\{ \begin{array}{l}
d(m+1)=2d(m)-\mu_1(m)-\mu_2(m), \vspace{.1truecm}\\
\mu_1(m+1)=d(m)-\mu_2(m), \vspace{.1truecm}\\
\mu_2(m+n)=d(m)-\mu_1(m),
\end{array} \right.
\end{equation*}
with initial conditions $d(0)=1$, $\mu_1(0)=0$ and $\mu_2(m)=0$,  for $m=0,\dotsc,n-1$.
The generating functions of the solution to this system of recurrence relations are given by:
\begin{eqnarray}
d(z)&=&1+2z+\dotsb+nz^{n-1}+\dfrac{(n+1)z^n}{1-z},\label{gf_n1-1}\\
\mu_1(z)&=&z+2z^2+\dotsb+(n-1)z^{n-1}+\dfrac{nz^n}{1-z},\notag\\
\mu_2(z)&=&\dfrac{z^n}{1-z}.\notag
\end{eqnarray}

Note that the degrees of $\phi_{+}^k$ grow linearly for $k=1,\dotsc,n-1$ and stabilize to $n+1$ for $k\geq n$.
This seems to be the first example of a birational map of $\deg=2$ with such behavior.




\section{Acknowledgements}

This research is supported by the DFG Collaborative Research Center TRR 109 ``Discretization in Geometry and Dynamics''.
The author would like to thank Matteo Petrera and Yuri Suris for their critical feedback on this manuscript.


\end{document}